\documentclass[11pt,a4paper]{article}
\pdfoutput=1

\usepackage{jheppub}
\usepackage{cleveref}

\usepackage{physics}
\usepackage{slashed}

\usepackage{microtype}
\usepackage{bm}
\usepackage{dsfont}

\usepackage{float}
\usepackage[export]{adjustbox}
\usepackage{pdfpages}

\def\ni{\noindent}

\makeatletter
\def\@fpheader{\vspace{0cm}}
\makeatother

\title{Tensors and Spinors in de Sitter Space}
\author{Ben Pethybridge, Vladimir Schaub}
\emailAdd{ben.pethybridge@kcl.ac.uk}
\emailAdd{vladimir.schaub@kcl.ac.uk}
\affiliation{Mathematics Department, King's College London, \\
The Strand, London,  WC2R 2LS, UK }
\abstract{We construct the Wightman function for symmetric traceless tensors and Dirac fermions in dS$_{d+1}$ in a coordinate and index free formalism using a $d+2$ dimensional ambient space. We expand the embedding space formalism to cover spinor and tensor fields in any even or odd dimension. Our goal is to furnish a self-contained toolkit for the study of fields of arbitrary spin in de Sitter, with applications to cosmological perturbation theory. The construction for spinors is shown in extensive detail. Concise expressions for the action of isometry generators on generic bulk fields, the 2-point function of bulk spinors, and a derivation of the uplift of the spinorial covariant derivative are included.}

\begin{document}
\maketitle
\newpage

\section{Introduction}

 In this paper we study efficient computational methods for spinorial and tensorial fields in pure de Sitter. The motivation for our work is based on two phenomenological facts: electrons exist, and space expands. From these simple remarks, comes a compelling incentive to study the behaviour of spin half fields in de Sitter spacetime, dS$_{d+1}$. During inflation, and our current era of dark-energy domination, the large scale structure of our universe seems to be accurately described, to leading approximation, by a conformally flat manifold with positive cosmological constant \cite{Spradlin:2001pw,Anninos:2012qw,Baumann:2009ds}. It was recognised to be computationally convenient to treat field theory from the perspective of an ambient space with an embedded dS slice since the earliest work on the subject by Dirac \cite{Dirac:1935zz,Dirac:1936fq}. Lifting fields to a higher dimensional space, to linearise the action of the isometry group and manifest the maximal symmetry, has been a technique developed thoroughly in the context of flat-space with conformal symmetry, as well as in euclidean AdS \cite{Weinberg:2010ws,Costa:2011wa,Costa:2011vf,Costa:2014kfa}. This formulation fostered new results in holography and greatly improved our conceptual understanding of QFT in curved spacetime \cite{Penedones:2010ue,Penedones:2016voo}. Well-honed methods have appeared to deal with arbitrary tensor representations \cite{Costa:2018mcg,Meltzer:2019nbs}, as well as spinors in odd dimension \cite{Nishida:2018opl}. Most importantly, it has become a commonplace tool to evaluate Witten diagrams in AdS$_{d+1}$. 

The ambient space picture arises naturally, given the realisation that the isometry group of dS$_{d+1}$ is  $SO(1,d+1)$. Consider the space $\mathbb{R}^{1,d+1}$, and its Lorentz-invariant subspaces. There are three classes of submanifold one can consider, defined by points $P^{A}\in \mathbb{R}^{1,d+1}$ with $P^{A}P_{A}= -1, 0, +1$, corresponding respectively to EAdS$_{d+1}$, the projective lightcone (i.e. the embedding of $CFT_d$), and dS$_{d+1}$. Each class of submanifold inherits a group action from the higher dimensional space, defined by the push forward of the isometry generators of $SO(1,d+1)$. This situation is visually appealing, as well as highly practical.
\begin{figure}[H]
	\centering
	\includegraphics[width=0.35\linewidth]{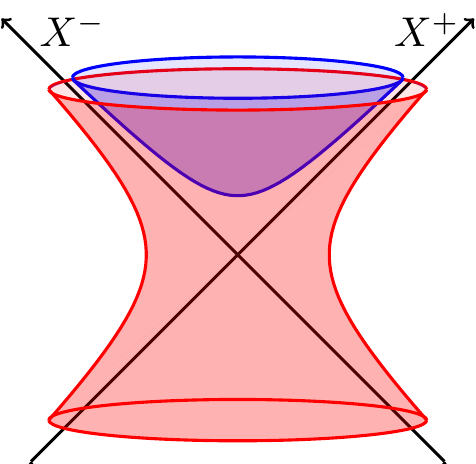}
	\caption{EAdS$_{2}$ (blue) and dS$_{2}$ (red) realised as Lorentz-invariant subset of $\mathbb{R}^{1,2}$, here represented with Lightcone axes $X^{\pm}=X^{0}\pm X^{2}$.}
\end{figure}
The benefits are numerous: for the most part one can avoid specific choices of a coordinate chart and differential operations become easy to write down, manipulate, and compute. In addition, the notion of the  boundary limit appears naturally by identifying points at the boundary of these spaces with points on the projective lightcone. One considers points on the submanifold $X^{A}X_{A}=1$, parameterised in terms of null vectors, $X^{A}=\lambda P^{A} + \frac{1}{\lambda}\ldots$ with $P^{A}P_{A}=0$. The limit $\lambda \rightarrow \infty$ gives insertions on the lightcone, which are identified with operators in a CFT$_{d}$. These operators are naturally classified by their representation of $SO(1,d+1)$, and therefore specified by a complex weight $\Delta$, and a representation of $SO(d)$. Examples of scalar operators of this type are discussed in \cite{Sengor:2019mbz,Sengor:2021zlc}.

The embedding picture for dS has received less attention. However, it has surfaced in calculations for bosonic quantities \cite{Sun:2020sgn,Xiao2014,Sleight:2021plv,Garidi:2003bg}, for both integer and half integer spin in the specific context of dS$_4$ \cite{Takook:2014paa}, and appeared as part of a wider effort to understand massless fields of general spin in maximally symmetric spacetimes \cite{Fronsdal:1978vb,Fang:1979hq,Huguet:2006fe,Faci:2009un}. It has also been used to consider the general properties of interacting two point functions in \cite{Bros:1995js,Bros:1994dn}. The goal of this paper is foremost to give a self-contained, dS oriented, entry point to these methods, with view toward future applications. Secondly we fill a gap in the literature, by constructing the ambient space approach for massive spinors in general dimensions and deriving their Wightman function. This approach easily generalises to the equivalent construction in (E)AdS, which has only been performed in odd dimension in the embedding \cite{Nishida:2018opl}, or in a specific coordinate system \cite{Henningson:1998cd}. We thus provide a convenient calculational toolkit for any integer or half integer spin, while also contributing to the literature on fermionic fields in de Sitter.

The topic of the propagator of a spin $\frac{1}{2}$ field in dS has received some attention previously. After an initial attempt within the ambient space formalism by Dirac \cite{Dirac:1935zz}, the calculation of the propagator for Dirac fermions in de Sitter has mostly been attempted using mode decomposition in the tetrad formalism. Work on the subject has been reported in \cite{Candelas1975,Koksma2009,Cotaescu2002,Cotaescu2018}, with which our results might be compared. Relevant calculations for spinors on euclidean maximally symmetric surfaces has been considered in \cite{Camporesi1992,Camporesi1996} and spin-half fields in dS have also appeared in the context of supersymmetric theories in \cite{Hertog:2019uhy}. Overall, the difference with (E)AdS \cite{Henningson:1998cd,Kawano:1999au} and CFT \cite{Iliesiu:2015qra,Costa:2011wa}, is striking. We present a thorough treatment of this subject as one of our goals. The main result of our work is a concise expression for the Wightman function of Dirac spinors in dS$_{d+1}$. We also give coordinate expressions for this 2-point function in planar and global coordinate systems.

The derivation of the Wightman function is required for further applications to interacting field theory. Crucially, they are used in the context of in-in formalism for perturbation theory \cite{Weinberg2005,Higuchi:2010xt,Gorbenko:2019rza}. Our computations are then of clear interest, given the recent inflation of published work studying QFT in dS. Previous work \cite{Sleight:2017fpc,Sleight:2019hfp,Sleight:2019mgd,Sleight:2020obc,Sleight:2021plv,DiPietro:2021sjt,maldacena2003,Bzowski2013,McFadden2010b,Pimentel2014,Anninos2014,Mata2013,Harlow:2011ke} focused on scalars, offers interesting insight relating QFT in dS and EAdS at the perturbative level. Meanwhile, numerous works have exploited the euclidean conformal symmetry of dS \cite{Hogervorst:2021uvp,Baumann2018,Baumann:2019oyu,Baumann:2020dch,Baumann:2020ksv,Arkani-Hamed:2018kmz,Arkani-Hamed:2017fdk,Goodhew:2021oqg,Jazayeri2021,Bonifacio2021}, these are also mostly concerned with bosonic quantities. It is our hope that the computation we present will help to extend this effort to fermionic fields, and deepen our understanding of QFT in dS in general.

\subsection*{Outline}
The plan of the paper is the following. We first review the construction of the bosonic 2-point function using the ambient space formalism in \cref{sec:scalar}, analogously to that of scalars \cite{Spradlin:2001pw} which is available in \cref{eq:Scalarprop}. This allows us to discuss the different charts one can use to cover the slice in \cref{def:planar,def:global}, as well as the analytical properties of the Wightman functions and their interpretation in perturbation theory in \cref{sec:inin}, and the late-time limit of operator insertions in \cref{sec:boundary}. In this construction we transpose the work of \cite{Costa:2014kfa} for spinning operators to the dS slice, with the extension to the Wightman function for the symmetric traceless tensor presented in \cref{app:STT}.  We then consider the uplift of Dirac fermions in \cref{sec:scalar}. We construct these using a method inspired by \cite{Weinberg:2010ws,Isono:2017grm}, matching the transformation law of spinors in the ambient space and on the slice which are written explicitly in \crefrange{eq:comphiD}{eq:comphiM}. We find how to constrain ambient spinors to obtain irreducible dS spinors, and showcase a formalism unifying this analysis in both even and odd dimensions with the final constraint given by \cref{eq:constraint}. We then perform our main computation, the Wightman function of Dirac spinors, by uplifting the Dirac equation and solving it in \cref{eq:spinW+,eq:spinW-}. Explicit expressions are given in this section for planar coordinates, and in \cref{app:Global} for global coordinates. We close with a discussion of our results in \cref{sec:discussion} and future directions. The appendices contain the aforementioned elements as well as a summary of our conventions in \cref{app:Conv} and a basic review of representation theoretic notions relevant to our work in \cref{app:repthy}. 

\section{Embedding Methods for Tensor Fields}

We start by reviewing the basic tools of the ambient space, developed thoroughly for EAdS$_{d+1}$ symmetric traceless tensors by \cite{Costa:2014kfa}, adapted to dS$_{d+1}$. Our aim is to set the logic and notation, and give a survey of the existing literature. The discussion of local coordinates and the scalar propagator is a classic topic, see for example \cite{Spradlin:2001pw,Anninos:2012qw}. The analytic structure of the Wightman function and its link to the in-in formalism is explained in its full generality and details in \cite{Weinberg2005}. 
    
 \subsection{Local coordinates on the dS slice} 
    
Consider dS$_{d+1}$ as a submanifold of $\mathbb{R}^{1,d+1}$.\footnote{We use indices $A,B,... =0,1,...d+1$ or $ A,B,... = +,-, 1,...d$ for light-cone coordinates. Lower case letters specify the ranges $a,b,..= 1,2,...d$ for latin characters at the beginning of the alphabet, $i,j,...=1,2,....d+1$ for those in the middle and, $\mu,\nu,...=0,1,...d $ as usual for greek indices. Contractions between indices will be performed with the Minkowski metric ($\eta_{\mu,\nu}$, $\eta_{AB}$) or the euclidean metric ($\delta_{ab}$,$\delta_{ij}$).} It corresponds to points $X^{A}$ satisfying 
\begin{equation}
	X^2 = - X^{+}X^{-}+X^{a}X_{a}= X_{d+1}^2+X^{\mu}X_{\mu}=-X_{0}^2+X^{i}X_{i}=1 \, .
\end{equation}
All our conventions can be found in \cref{app:Conv}. In what follows $X$ will generically denote an ambient space vector field satisfying such a constraint.  We obtain a dS-foliation of the spacelike region of the ambient space by multiplying $X^{A}$ by a real number $\ell>0$, however we work exclusively in the regime with fixed unit de Sitter length, equivalent to $\ell =1$. An example of a parametrisation of the de Sitter slice is given by conformally flat (planar) coordinates, the analogue of Poincar\'e coordinates in AdS  
\begin{equation}
	X^{A}= (X^{+},X^{-},X^{a})=\frac{1}{\eta}(1,x^2-\eta^2,x^{a}) \, \label{def:planar}.
\end{equation}
Note $x^a \in \mathbb{R}$ and we choose $\eta<0$, where $\eta$ increases from $-\infty$ towards $0^{+}$. This patch covers the causal future of an observer sitting at the origin in the far past, and therefore includes the late time slice. These coordinates are the simplest one can use, and usually most convenient for explicit calculations. Another interesting choice are global coordinates 
\begin{equation}
	X^{A}= (X^{0},X^{i})=(\sinh t, \omega^i\cosh t ) \, \label{def:global},
\end{equation}
Where $\omega ^i$ parameterise a spacelike $S_d$ and $ \omega^i \omega_i = 1$. These coordinates cover the entire slice.

The formalism of this paper avoids the necessity of choosing a coordinate patch. However, we make use of the planar patch to make our construction explicit and prove some coordinate independent results in \cref{sec:Spinor}, while \cref{app:Global} contains the specialisation of our methods to global coordinates.

More generally, local coordinates $x^{\mu}$ chart coordinate patches on the slice, providing a map $X^{A}(x^{\mu})$. From this map, we can unfold the machinery of differential geometry of a submanifold, albeit greatly simplified since the ambient space is flat \cite{bruhat:1982}. One can define frame-fields $e_{\mu}{}^{A} \equiv \pdv{X^{A}}{x^{\mu}}$, which defines the push-forward of ambient tensors to the slice. For instance, we can recover the planar metric
\begin{equation}
	ds^2 = \frac{-d\eta^2+dx_a dx^a}{\eta^2}= \frac{dx_{\mu}dx^{\mu}}{x_0^2} \, .
\end{equation}
 Note that, to simplify notation, the lower case indices of the slice are all $SO(1,d)$ tangent space indices, and are therefore contracted with the Minkowski metric. To recover the spacetime quantities these should be contracted with the ordinary tetrad (as opposed to the ``frame field'' we have just defined). Though the use of specific coordinates is practically necessary to perform computations intrinsically, the main advantage of the embedding picture is that for almost all computations, we do not need to choose a specific parameterisation. One can work only with embedding objects, which are in one-to-one correspondence with local objects on the slice. 

\subsection{Tensor fields and differential operators} 
We illustrate the construction using tensor fields. First, notice that $e_{\mu}{}^{A}X_{A}=0$, hence all longitudinal components of a tensor have zero projection on the dS slice. This redundancy can be fixed by considering only transverse tensors, effectively, we choose a gauge-fixing. From this, we define the uplift of tensorial operators on dS as transverse tensor fields in the ambient space. 

For example, a Symmetric Traceless Tensor (STT) operator $T_{A_1 \ldots A_l}(X)$ in the ambient space is a dS tensor of same rank and symmetry properties, provided
\begin{equation*}
	X^{A_i}T_{A_1 \ldots A_l}(X) =0 \, .
\end{equation*}
%.
The uplift of the induced metric is important, as 
\begin{equation}
	G_{AB}=\eta_{AB}-\frac{X_{A}X_{B}}{X^2}=\eta_{AB}-X_{A}X_{B} \, ,
	\label{def:projection}
\end{equation}
defines a projector to the slice. dS objects are formed by contracting indices with $G_{AB}$. Note the change of sign in \cref{def:projection} with respect to the result in EAdS \cite{Costa:2014kfa}. Contracting all indices with $G_{AB}$ ensures that we only evaluate fields on the slice on which they are defined. For example the covariant derivative
\begin{equation}
	\nabla_{A}= G_{A}{}^{B}\pdv{}{X^{B}}= \partial_{A}-X_{A}X^{B}\partial_{B} \, ,
	\label{eq:CovDiv}
\end{equation}
transforms only between objects defined in dS. It acts on a generic tensor field $T_{A \ldots}$ as
\begin{equation}
	\nabla_{A}T_{B\ldots }=G_{A}{}^{A'}G_{B}{}^{B'}\ldots \pdv{T_{B'\ldots}}{X^{A'}} \, .
\end{equation}
%.

In practice, indices can and should be avoided when possible, and an index-free formalism is favoured here. This is done by considering STT operators as scalar, homogeneous polynomials in a dummy transverse and null vector variable $W^{A}$. i.e. given $T_{A_1 \ldots A_l}(X)$, we take $T(X,W)\equiv W^{A_1}\ldots W^{A_{l}}T_{A_1 \ldots A_l}(X)$, with polarisation vector $W^{A}$ such that $W\cdot X = W^2 = 0$.\footnote{One can extend the index-free formalism to mixed-symmetry tensor in a straightforward way following \cite{Costa:2014rya}, by using multiple polarisation vectors and imposing the symmetry of the Young tableau, by hand or using grassmanian variables.} The first condition ensures that $W^{A}$ is a dS vector, fixing the redundancy; the second simplifies all traces. One can check that together the conditions on $W^A$ imply that for any choice of parameterisation $x^{\mu}$, $W^{A}\equiv w^{\mu}e_{\mu}{}^{A}$, with $w^{\mu}w^{\nu}g_{\mu\nu}=0$. In a sense, polarisation takes care of the projection, if $t_{\mu_1 \ldots \mu_l}$ is the push forward of $T_{A_1 \ldots A_l}$, then $T(X(x),W(x,w))=t(x,w)$. Having contracted all indices, it can be necessary to free them once again. This is done by using a vector differential operator of homogeneous degree $-1$ in $W$, whose image only contains STT tensors, and preserves the constraint $X\cdot W = W^2 =0$. These choices uniquely fix it to be the uplift of the Todorov operator, $K_A$ defined by
\begin{equation}\label{eq:KA}
\begin{aligned}
	K_A \equiv \left(\frac{d-1}{2}+W\cdot \pdv{}{W}\right)\left(\pdv{}{W^{A}}-X_A X\cdot \pdv{}{W}\right)-\frac{W_{A}}{2}\left(\frac{\partial^2}{\partial W \cdot \partial W}- \left(X\cdot \pdv{}{W}\right)^2\right) \, .
	\end{aligned}
\end{equation}
The Todorov operator has been used in the CFT literature\cite{Costa:2011wa}, and also appears in conformal geometry as the Thomas operator \cite{Curry:2014yoa}.\footnote{One should note that subsequent freeing and contraction of indices returns the same object up to a spin-dependent constant. This can be checked in an example 
\begin{align*}
	W\cdot K (W\cdot V)^{J} = J\left(\frac{d+1}{2}+J-2\right) (W\cdot V)^{J} \, .
\end{align*}}
In the index-free notation, the covariant derivative is slightly modified, as an explicit computation shows 
\begin{equation}
	\nabla_{A}T(X,W)=\left(\pdv{}{X^{A}}-X_{A}X\cdot \pdv{X}- W_{A}X\cdot\pdv{}{W}\right)T(X,W) \, .
\end{equation}
\subsection{Propagators \label{sec:scalar}}

We now consider the 2-point function of free bulk scalar fields $\phi$, $\Pi(X,Y)=\expval{\phi(X) \phi(Y)}$ with respect to a de Sitter invariant vacuum state. We choose the Bunch-Davies vacuum through appropriate boundary conditions. We derive the Wightman function, which obeys the homogeneous equations of motion. Lorentz invariance of the ambient space forces this function to depend only on the geodesic distance between points $u=(X-Y)^2$. A slightly more convenient choice of variable is the de Sitter invariant distance $z=1-\frac{u}{4}=\frac{1}{2}(1+X\cdot Y)$. Therefore $z=1$, $z >1 $, $z<1$ implies the points are seperated by null, timelike or spacelike geodesics respectively, $z = 0$ implies X is null with respect to the antipodal point of Y \cite{Spradlin:2001pw,Anninos:2012qw}. We rewrite the Casimir eigenvalue problem as an ODE for $g(z)=\Pi(X,Y)$, where $\mathcal{C}$ defines the Casimir operator of the de Sitter group
 \begin{equation}
      (\mathcal{C} - \Lambda)\Pi(X,Y) = 0 \label{eq:casimire}\, .
 \end{equation}
The action of $\mathcal{C}$ and the eigenvalue $\Lambda$ is fixed by the representation of $SO(1,d+1)$.  The uniqueness of the Laplacian as a second order differential operator commuting with the isometry of the de Sitter slice, implies the Klein-Gordon and Casimir equation are interchangeable. This is true up to a slight change in the eigenvalue for spinning fields \cite{Sun:2020sgn}. 

The de Sitter isometries are linearly expressed in the ambient space, where the Casimir operator takes the form
  \begin{equation}
      \mathcal{C} = - \frac{1}{2} L_{AB}L^{AB} \, . 
  \end{equation}
 The operators $L_{AB}$ generate the Lorentz group in the ambient space, through some differential realisation $\mathcal{L}_{AB}$ and possible spin parts. Throughout this paper we make use of the appropriate realisation for the various representations.  For the scalar case the eigenvalue is fixed by the conformal weight and dimension of the space $\Lambda = \Delta(d-\Delta )$ and we only need 
 \begin{equation}
 \mathcal{L}_{AB} = X_A\partial_B - X_B \partial_A\, .
 \end{equation}
 The Wightman function must solve the equation of motion given in \cref{eq:casimire}. The hypergeometric differential equation \cite{macdonald2013hypergeometric} is found by changing variable to $z$,
  \begin{equation}
     (1-z) z g''(z)+\frac{d+1}{2}\left(1-2z\right)  g'(z)-\Delta (d-\Delta ) g(z)=0 \, ;
  \end{equation}
  Generic solutions are given by linear combinations
  \begin{equation*}
	g(u) = \kappa \, _2F_1\left(d-\Delta,\Delta ;\frac{d+1}{2};z\right)+\tilde{\kappa} \, _2F_1\left(d-\Delta,\Delta ;\frac{d+1}{2};1-z\right) \, .
	\label{eq:scalarsol}
\end{equation*}
We must consider the analytic structure of this function. First of all, note that both functions diverge at the coincident limit, while the second one diverges also at antipodal points $X=-Y$. This latter singularity is screened behind the dS horizon. We consider Wightman functions with singularities only at coincident points, thus setting $\tilde{\kappa}=0$. This choice of boundary condition selects the Bunch-Davies vacuum from the continuum of possible vacua, often referred to as $\alpha$ vacua in the literature \cite{Bousso:2001mw,Allen:1985ux,Sasaki:1994yt}. Finally, we must fix the normalisation. Noting that in the coincident limit, the Wightman function is blind to the curvature of space, we match the normalisation of the flat-space result \cite{Strominger:2001pn} 
\begin{equation}\label{eq:Scalarprop}
    g(u) = \frac{\Gamma (\Delta ) \Gamma (d-\Delta)}{(4 \pi )^{d+1/2} \Gamma \left(\frac{d+1}{2}\right)} \, _2F_1\left(d-\Delta,\Delta ;\frac{d+1}{2};z\right).
\end{equation} 
We refer to the Wightman function obtained from this choice of boundary condition as the Hadamard form of the 2-point function. 

 In euclidean signature, this would be the final answer. However, rotating to lorentzian time, the hypergeometric function develops a branch-cut along time-like separations. We must specify how to evaluate this, in a way that specifies the time-ordering of the different points. This subtlety is slightly obscured in this formalism for which $\Pi(X,Y)=\Pi(Y,X)$. The $i\epsilon$ prescription must be brought back by hand when needed, usually at the coordinate level. We describe this process in \cref{sec:inin}, where we briefly recapitulate the in-in formalism for perturbation theory in de Sitter \cite{Weinberg2005,Higuchi:2010xt}. 

For the purposes of review we reserve the case of the STT propagator for \cref{app:STT}, the equivalent calculation in EAdS is performed in \cite{Costa:2014kfa}. For tensorial correlation functions, one must sum over all allowed structures, which involves a choice of basis. For example in the case of the spin-$1$ Wightman function, we have equivalent possible bases
\begin{equation}
\begin{aligned}
	\Pi(X,W_1; Y,W_2)&= (W_1\cdot W_2)g_0(u)+ ((W_1\cdot Y)( W_2  \cdot X))g_1(u) \\
	&=  (W_1\cdot W_2)f_0(u)+((W_1\cdot \nabla_{X})( W_2 \cdot \nabla_{Y}))f_1(u).  
	\label{def:Spin1bases}
\end{aligned}
\end{equation}
Which can be generalised to higher spin. When we perform the computation for spinors, we will similarly find a sum of allowed structures multiplied by scalar functions.

\subsection{The perturbative prescription \label{sec:inin}}

The observables most regularly calculated in de Sitter are insertions of operators on a single time slice with respect to the Bunch-Davies vacuum. We assume the theory approaches the free theory far in the past, where the fields act on the Bunch-Davies vacuum as in Minkowski spacetime. Splitting the Hamiltonian between the quadratic part, $H_0$, and the interacting Hamiltonian, $H_{int}$, we seek time evolution operators $U(t,t_0)$ which satisfy
\begin{equation}
    \mathcal{Q}(t) = U^{\dag}(t,t_0)\mathcal{Q}_I(t)U(t,t_0) \, .
\end{equation}
Where $\mathcal{Q}_I(t)$ is an operator including only the field insertions which evolve with respect to $H_0$ and therefore propagate as free fields. The operator $U(t,t_0)$ solves the same ODE as in the Minkowski case, and performing the similar calculation for the inverse leaves us with
\begin{align}
    U(t,t_0) &= T\left\{\exp\left[-i\int^t_{-\infty^+}dt_r H_I(t_r)\right]\right\}\\
    U^\dag(t,t_0) &= \bar{T}\left\{\exp\left[i\int^t_{-\infty^-}dt_l H_I(t_l)\right]\right\}\, .
\end{align}

$H_I$ is the evolution, with respect to $H_0$ of $H_{int}$ and the $\bar{T}$ implies the expansion should be anti-time ordered, with the insertions of operators placed in order of increasing $t$ from left to right. We have now taken the limit and prescription $t_0 \rightarrow -\infty(1\pm i\epsilon) = -\infty^\pm$, for the time evolution operator and inverse respectively. This implies the following $i\epsilon$ prescription for $t$
\begin{equation}
    t_l \rightarrow t_l(1+i\epsilon)\, ,\hspace{2cm}  t_r \rightarrow t_r(1-i\epsilon)\, . 
\end{equation}
From the expressions above, Wick contraction will require three different propagators, all of which may be derived from the Wightman function. In planar coordinates, we may rewrite the prescription in terms of the invariant length $z$ and $\alpha  = \text{sgn}(\eta_l-\eta_r)$. Labelling contractions of fields with $l$/$r$ for inclusion in  $T$ and $\bar{T}$ respectively, we require the propagators
\begin{align}
    \Pi_{ll}(z) &= g(z+i\epsilon)\label{def:ll}\\
    \Pi_{rr}(z) &= g(z-i\epsilon)\\
    \Pi_{lr}(z) &= g(z-i \alpha \epsilon)\label{def:lr}
\end{align}
In general the prescription for $z$ will be parameterisation dependent. A coordinate independent formulation seems slightly out of reach, although we note that the same prescription for $z$ is true in global coordinates. This formalism is reviewed and used in many places, among which notable examples include \cite{DiPietro:2021sjt,Adshead2009,Gorbenko:2019rza,Sleight:2021plv}. 

\subsection{Boundary limit \label{sec:boundary}}

From the embedding picture, the discussion of the boundary limit of operator insertion is greatly streamlined. The starting point is to choose a specific limiting geometry, by fixing a gauge choice for the projective null vector $P^{A}$ corresponding to boundary insertions. One can parameterise a generic point on the dS slice $X$ using a pair of null vectors, $P$ and $Q$ such that $ P^2 = Q^2= 1- 2 P \cdot Q =0 $ and a single variable $\lambda$ 
\begin{align}
    X^{A}&=\lambda P^{A}+\frac{1}{\lambda}Q^{A}
\end{align}
The boundary limit is achieved by taking $\lambda \rightarrow \infty$, which maps projectively the insertions of fields in the dS slice to those at points $P^{A}$ on the lightcone. Given a specific coordinate system, there is usually an obvious parameterisation of this type. For example, for planar and global coordinates we can write
\begin{align}
    \left(X^{+},X^{-},X^{a}\right) &=\left(\frac{1}{\eta},\frac{x^2-\eta^2}{\eta},\frac{x^{a}}{\eta}\right)=\frac{1}{\eta}\left(1, x^2,x^{a}\right) + \ldots  \, , \\
      \left(X^{0},X^{i}\right) &=\left(\sinh(t),\cosh(t)\omega^{i}\right)=\frac{e^{t}}{2}\left(1,w^{i} \right) + \ldots  \, .
\end{align}
Order by order in the expansion of the correlator, one can identify the correlator of the bulk field and some boundary insertions. At leading order, one obtains the bulk-to-boundary correlator for the boundary primary operator. For example, consider the Wightman function found previously, 
\begin{equation}
\begin{aligned}
    \lim_{\lambda \rightarrow \infty} \Pi(X=\lambda P + \ldots,Y) &= \lambda^{-\Delta}\frac{1}{(-2P\cdot Y)^{\Delta}}\frac{4^{\Delta } \Gamma (\Delta ) \Gamma (d-2 \Delta )}{(4 \pi )^{\frac{d+1}{2}} \Gamma \left(\frac{d+1}{2}-\Delta \right)} \\
    &+ \lambda^{-(d-\Delta)}\frac{1}{(-2P\cdot Y)^{d-\Delta}}\frac{4^{d-\Delta}\Gamma (d-\Delta ) \Gamma (2 \Delta-d)}{(4 \pi )^{\frac{d+1}{2}} \Gamma \left(\frac{1-d}{2}+\Delta\right)}
\end{aligned}\end{equation}
The pair of leading terms can be identified with the bulk to boundary correlator of the scalar field $\phi$ a conformal primary field $\mathcal{O}_{\Delta}(\lambda P)=\lambda^{-\Delta}\mathcal{O}_{\Delta}(P)$ and its shadow dual, $\mathcal{O}_{d-\Delta}(P)$.

\section{Spinors in dS$_{d+1}$ \label{sec:Spinor}}
 We turn to a systematic treatment of spinors. We begin with ambient Spinor fields, and constructively show, using the methods of \cite{Weinberg:2010ws,Isono:2017grm}, how to constrain them to obtain irreducible spinors of the dS slice in \cref{eq:constraint}. As a by-product of this analysis, we write the commutation relations of symmetry generators with fields of generic spin in the bulk of dS in \crefrange{eq:comphiD}{eq:comphiM}. We introduce an index-free notation for spinors, and showcase these tools by computing the propagator of Dirac spinors in dS$_{d+1}$ given by \cref{eq:spinW+,eq:spinW-}. We finally discuss the late-time limit of the propagator given in \cref{eq:spinWblim}. Odd and even dimension are initially treated separately, though the final result for Dirac spinors is shown to be equivalent. The case of spinors in odd EAdS was covered in \cite{Nishida:2018opl}, our added value lies in giving an explicit construction which generalises both to other coordinates and to even dimensions. 
   
  \subsection{Constraint and transformation law} 
    
Embedding the dS slice leads to a straightforward realisation of its isometries, mapping them to Lorentz transformations. Defining $\Sigma_{AB} \equiv \frac{1}{4} [ \Gamma_A,\Gamma_B ]$ and $ \mathcal{L}_{AB} \equiv P_A \pdv{}{P^{B}}-P_B \pdv{}{P^{A}}$ the transformations act linearly on the ambient space spinor field $\Psi$ via
\begin{align}
    [L_{AB}, \Psi(P) ] &=- (\mathcal{L}_{AB}+\Sigma_{AB})\Psi(P) \label{eqn:LABPhiAmb}\, , \\
    [L_{AB},L^{CD}]&=-\eta_{A}{}^{C}L_{B}{}^{D}+\eta_{B}{}^{C}L_{A}{}^{D}+\eta_{A}{}^{D}L_{B}{}^{C}-\eta_{B}{}^{D}L_{A}{}^{C} \, , \\
    &=-4 \eta_{[A}{}^{[C}L_{B]}{}^{D]} \, ,
\end{align}

 $L_{AB}, \mathcal{L}_{AB}$ and $\Sigma_{AB}$ obey the same commutation relations. In what follows we consider $\Psi$, a Dirac spinor representation of $SO(1,d+1)$; and $\overline{\Psi}$, its conjugate. As we did previously for tensors, we will \textit{define} a dS spinor as a constrained object which lives in the ambient space. The Lorentz invariant constraint will make this object irreducible, in the same way that transversality did, and ensure that it contains degrees of freedom corresponding exactly as those of a dS spinor. We perform our construction in a specific parameterisation of the dS slicing $X^2 = 1$, but the results and constraint are coordinate independent. The transformation law \eqref{eqn:LABPhiAmb} maps to a local realisation of the isometries acting on components of $\Psi$ on the slice; matching the local realisation to the intrinsic calculation implies a constraint on $\Psi$ which corresponds to an irreducible field $\psi$ in dS. For the planar parameterisation the algebra is most naturally written using 
\begin{align}
	D &= 2L_{+,-} \, , \\
	P_a &= 2L_{-,a} \, , \\
	K_a &= 2L_{a,+} \, , \\
	M_{ab} &= L_{ab} \, .
\end{align}
When written explicitly in planar coordinates, these operators act on a scalar field $\phi$ as Killing vectors specified by
\begin{align}
	\hat{Q}_{D}\phi &= x^{\mu}\pdv{}{x^{\mu}} \phi \, ,\\
	\hat{Q}_{P_a}\phi &= \pdv{}{x^{a}}\phi \, ,\\
		\hat{Q}_{K_a}\phi &= \left(2 x_{a} x^{\mu}\pdv{}{x^{\mu}}-x^{\mu}x_{\mu}\pdv{}{x^{a}}\right)\phi\, , \\
	\hat{Q}_{M_{ab}} &=\left( x_{b}\pdv{}{x^{a}}-x_{a}\pdv{}{x^{b}}\right)\phi\, .
\end{align}
The intrinsic action on other dS representations is fixed by the addition of a spin part to the action of the operators. For a field transforming in a given representation of $SO(1,d)$, we add a contribution from the spin matrix, for spinor that is $\tilde{\Sigma}_{\mu \nu} = \frac{1}{4}[\gamma_\mu,\gamma_\nu]$, to the action of $M_{ab}$. We then use the Jacobi identity to fix the spin-part of all the remaining generators in terms of the spin matrix $\tilde{\Sigma}_{\mu\nu}$. Note that the indices used here are the $SO(1,d)$ tangent space indices as described in \cref{app:Conv}, as such $\{\gamma_\mu,\gamma_\nu\} = 2 \eta_{\mu\nu}$. This analysis provides the general form of the isometry generators acting on spinning dS bulk fields $\psi$ in planar coordinates,
\begin{align}
	[D,\psi(x^{\mu})]&=x^{\mu}\pdv{}{x^{\mu}}\psi \label{eqn:LABPhiinti} \, ,\\
	[P_{a},\psi(x^{\mu})]&=\pdv{}{x^{a}} \psi \, , \\
	[K_{a},\psi(x^{\mu})] &= \left(2 x_{a} x^{\mu}\pdv{}{x^{\mu}}-x^{\mu}x_{\mu}\pdv{}{x^{a}}+2\tilde{\Sigma}_{a\mu}x^{\mu}\right)\psi \, ,\label{eq:commutK} \\
	[M_{ab},\psi(x^{\mu})]&=-\left(x_{a}\pdv{}{x^{b}}-x_{b}\pdv{}{x^{a}}+\tilde{\Sigma}_{ab} \right)\psi \, \label{eqn:LABPhiintf}.
\end{align}
Note that $\tilde{\Sigma}_{ab}$ needs not be the $ab$ component of $\Sigma_{AB}$, though it is often the case.  We show the analogous calculation for the related example of global coordinates in \cref{app:Global}. To our knowledge, these commutations relations, with spin contributions, are not present in the literature. Though this appears superficially similar to the transformation law of a field of weight $0$ under the conformal group, one should be mindful of the index range. Only in the late time limit $\lim_{\eta \rightarrow 0^{+}}\psi(\eta,x)= (-\eta)^{\Delta}\psi(x)$, do we recover the usual form of the Conformal algebra acting on primary fields of weight $\Delta$.

We can now compare how the elements of $\Psi$ transform in the ambient space, \cref{eqn:LABPhiAmb} with the transformation in the planar patch, \crefrange{eqn:LABPhiinti}{eqn:LABPhiintf}, to isolate a quantity transforming like a spinor of dS. Though an explicit choice is made, we stress that the results are not unique to the planar parameterisation, precisely because of the homogeneity of dS. As further proof of this, in the \cref{app:Global} we perform the analogous analysis for global coordinates.
    
\subsubsection{Uplifting spinors of dS$_{2n+1}$}
    
The embedding space is now even dimensional, and we decompose the ambient space spinor as a direct sum of two spinors in $Spin(1,2n)$,
\begin{align}\label{eq:SpinorRep}
	\Psi = \begin{pmatrix}
		\chi \\
		\rho \end{pmatrix}\, .
\end{align}
Associated to the dS slice we have matrices $\gamma_\mu$, from which we construct those of the ambient space 
\begin{align}\label{eq:Gammai}
    \Gamma_0 &=  \begin{pmatrix} 0 && 1 \\ -1 && 0 \end{pmatrix}\otimes\mathds{1}  \, , \\
    \Gamma_a &=\begin{pmatrix} 1 && 0 \\ 0 && -1 \end{pmatrix}\otimes \gamma_0\gamma_a  \, ,\\ 
     \Gamma_{d+1} &= \begin{pmatrix} 0 && 1 \\ 1 && 0 \end{pmatrix} \otimes \mathds{1}\, ,\\
     \Gamma_\star &=   \begin{pmatrix} 1 && 0 \\ 0 && -1 \end{pmatrix}\otimes-i\gamma_0\label{eq:Gammaf} \, .
\end{align}
From these choices, one may compare $[L_{AB},\chi]$ in the ambient space, \cref{eqn:LABPhiAmb}, with that defined by \crefrange{eqn:LABPhiinti}{eqn:LABPhiintf}. Since the $-+$ terms give an inhomogeneous contribution to the dilation, we consider a rescaled field $\sqrt{-\eta}\chi$, simplifying the action on the fields to
\begin{align}
	[D,\sqrt{-\eta}\chi(x^{\mu})]&=x^{\mu}\pdv{}{x^{\mu}}\left(\sqrt{-\eta}\chi\right) \, ,\label{eq:comphiD} \\
	[P_{a},\sqrt{-\eta}\chi(x^{\mu})]&=\pdv{}{x^{a}} \left(\sqrt{-\eta}\chi\right) \, , \\
	[K_{a},\sqrt{-\eta}\chi(x^{\mu})] &= \left(2 x_{a} x^{\rho}\pdv{}{x^{\rho}}-x^{\rho}x_{\rho}\pdv{}{x^{a}}\right)\left(\sqrt{-\eta}\chi\right)-\sqrt{-\eta}\left(x_a \chi-\gamma_a\gamma_0 \rho\right) \label{eqn:Ketaphi} \, ,\\
	[M_{ab},\sqrt{-\eta}\chi(x^{\mu})]&=-\left(x_{a}\pdv{}{x^{b}}-x_{b}\pdv{}{x^{a}}+\frac{1}{4}[\gamma_a,\gamma_b]\right)\left(\sqrt{-\eta}\chi\right) \,\label{eq:comphiM} .
\end{align}
Clearly $\sqrt{-\eta}\chi$ has the correct transformation law under translations, rotations, and the dilation. The special conformal transformations are more problematic, with an inhomogeneous contribution from the second spinor $\rho$ in \cref{eqn:Ketaphi}. We regain the coordinate transformation given in \cref{eq:commutK} by setting $\rho = -\gamma_0 \gamma_\mu x^{\mu}\chi$. This can be rewritten as a Lorentz-covariant constraint on the ambient spinor $\Psi$, 
 \begin{align}
 	\Gamma_A X^{A} \Psi= \Psi \, .
 	\label{eq:constraint}
 \end{align}
At this point, we stress the analogy with the tensorial case. We showed that generic ambient spinors do not define dS spinors. However, there is a specific class of (Lorentz-covariantly) constrained spinors in the embedding which are in one-to-one correspondence with spinors on the slice. Hence, we can define the uplift of dS spinors as precisely those constrained spinors, in total correspondence with what we saw for tensors. One can also note that this constraint is the only possibility, since $(X\cdot \Gamma)^2=1$, and the spinor with eigenvalue $-1$ is related to the one we consider by multiplication with $\Gamma_{\star}$. 

When this argument is reversed, we see that any spinor of the form $\left(\Gamma_A X^{A}+1\right)\Psi$, with $\Psi$ unconstrained, has a top component which transforms like $\frac{1}{\sqrt{-\eta}}$ times a dS spinor. In analogy to the tensorial case, we can work with unconstrained spinors like \eqref{eq:SpinorRep}, and contract them with a dummy constrained conjugate polarisation spinor $\overline{S}$, which incorporates the projection prefactor. This allows us to work with scalar fields $\bm{\Psi}(X,\overline{S})\equiv \overline{S}\Psi(X)$, instead of spinors. On the slice, we also work with the scalar $\overline{s}\psi(\eta,x)$, allowing us to avoid the use of spinor indices. The constraint is then transferred to the polarisation, $\overline{S}$, which also takes care of the projection, as in the tensorial case. The dummy spinor in the ambient space and in the slice are related directly, 
\begin{equation}
     \overline{S}\Psi= \overline{s}\psi \, .
\end{equation}
We can use the constraint to derive the ambient space $\overline{S}$ in terms of the intrisic $\overline {s}$, taking the complex conjugate we recover the projection to conjugate spinors. The spinor polarisations are given by : 
\begin{align}
    X^A \Gamma_A S = -S  \hspace{1cm}&\implies\hspace{1cm} S = \begin{pmatrix} \gamma_0 \\ -\slashed{x} \end{pmatrix}\frac{s}{\sqrt{-\eta}} \, , \\
     \overline{S}X^A \Gamma_A = \overline{S}\hspace{1cm}&\implies\hspace{1cm} \overline{S} = \frac{\overline{s}}{\sqrt{-\eta}}\begin{pmatrix} -\slashed{x}\gamma_{0} && 1 \end{pmatrix}  \, .
\end{align}
We define dS spinors, in analogy to the tensor case, as homogeneous scalar polynomials in the (commuting) polarisation variables
\begin{equation}
    \bm{\Psi}(X,\lambda \bar{S})=\lambda \bm{\Psi}(X, \bar{S}) \, ,
\end{equation}
with obvious generalisation to multiple spinor indices. It is of course possible to work directly with constrained spinors, with free indices. Then, one must write out spinor-structures which manifestly obey the constraint. Switching to the index free notation only makes manipulation simpler, one can simply write all non-vanishing scalar objects satisfying the constraint on $S$ and the homogeneity requirement. 

The explicit expressions we gave for $S$ are specific to the choice of parameterisation of dS, one can translate these back into other choices, though this can be a lengthy endeavour. We showcase the computation for global coordinates in \cref{app:Global}. In this work we preferentially compute results directly in the embedding, and only evaluate in a specific parameterisation, if necessary, as a final step. 

\subsubsection{The case of dS$_{2n}$}
    
In even dimensions the spin group admits irreducible Weyl representations, this more intricate case requires a slightly more subtle approach. We start by discussing the transformation induced by the Lorentz group of the ambient space, and derive the uplift of the dS Weyl-spinors. We then use this result to reconstruct the full Dirac spinor in even dimensions. 
    
The ambient space is odd dimensional. So naively, the Dirac spinor representations are the same on the slice and in the embedding, and we could be tempted to try to use unconstrained ambient objects. However, from the previous computation it is clear that such an object does not transform as it should on the slice. One may expect from the case of even ambient space that the constrained Dirac spinor maps to local irreducible spinors of the slice, which are the Weyl spinors in even dS. Inspired by these remarks, we consider the chiral representation of the gamma matrices on the slice 
\begin{align}
     \gamma_\mu &=\begin{pmatrix} 0 && \sigma_\mu \\ \overline{\sigma}_\mu && 0 \end{pmatrix} \, , \\ 
    \sigma_\mu &= (\mathds{1}, \sigma_a) = \sigma_\mu^\dagger \, , \\
    \overline{\sigma}_\mu &= (-\mathds{1}, \sigma_a) = \overline{\sigma}_\mu^\dagger \, .
\end{align}
Consider the basis of ambient gamma matrices in \crefrange{eq:Gammai}{eq:Gammaf}, with the lower slice $\gamma$-matrices replaced by chiral $\sigma$-matrices
\begin{align}
    \varGamma_0 &=  \begin{pmatrix} 0 && 1 \\ -1 && 0 \end{pmatrix} \otimes \mathds{1}\, , \\
    \varGamma_a &=  \begin{pmatrix} 1 && 0 \\ 0 && -1 \end{pmatrix}\otimes\sigma_a \, , \\ 
     \varGamma_{d+1} &= \begin{pmatrix} 0 && 1 \\ 1 && 0 \end{pmatrix}\otimes\mathds{1} \, .
\end{align}
Many of the results of the previous section can be reused. For example, the adjustment of the action of $D$ and $P$ are identical. Rotations indeed involve $\Sigma_{ab}=\frac{1}{4}[\sigma_a,\sigma_b]$ which is identical for left and right handed Weyl spinors. For the boosts we encounter $\Sigma_{a\mu}x^{\mu}=\frac{1}{2}\left(\sigma_a\left(-\alpha\eta\mathds{1}a +\sigma_b x^{b} \right)-x_a\right)$, with $\alpha = \pm 1$ corresponding to LH-spinors and RH-spinors respectively. It follows that an ambient spinor of the form 
\begin{equation}
    \sqrt{-\eta}\begin{pmatrix} \chi \\ \left(\alpha \eta\mathds{1} - x^{a}\sigma_a \right)\chi \end{pmatrix} \, ,
\end{equation}
encodes a chiral spinor in its top component,  with $\alpha = \pm 1$ defining the handedness. Such a spinor has an eigenvalue equation of the form $X^{A}\varGamma_A \Psi_{\alpha} = \alpha \Psi_{\alpha}$. The key difference is the absence of $\varGamma_{\star}$ to exchange $\alpha$, implying the sign is now a meaningful distinction between inequivalent representations. Since any such spinor can be written as $\left(X^{A}\varGamma_A + \alpha \right)\Psi$ with $\Psi$ unconstrained, we can proceed as in the odd case to define a polarisation spinor $S$, with eigenvalue equation $\overline{S} X\cdot \varGamma = \alpha \overline{S}$.\footnote{This chiral picture will be useful to those interested in supersymmetry in de Sitter and chiral interactions, but for many purposes it is easier to work with the reducible Dirac spinor.}

To discuss the uplift of the Dirac spinor, it is convenient to adopt the terminology of \cite{Trautman:1992}. In even dimensions, we have two Weyl spinors which together form a Dirac spinor. By analogy, in odd dimension, we call the irreducible representation, of Dirac type, a Pauli spinor, and an $SU(2)$ doublet of Pauli spinors a Cartan spinor. We have showed that in even dimensional ambient space, a constrained Dirac spinor encodes a Pauli spinor on the slice. On the other hand the constrained Pauli spinor encodes the Weyl spinors when the ambient space has odd dimension. It follows naturally that we can build a constrained Cartan spinor which uplifts a Dirac spinor. Effectively, we recompose the uplift of the Dirac spinor as a sum of its Weyl parts $\Psi' = \Psi_{+}\oplus \Psi_{-}$. We pick Gamma matrices given by $\Gamma'_{A}=\varGamma_{A}\otimes\big(\begin{smallmatrix}
  1 & 0\\
  0 & -1
\end{smallmatrix}\big)$, such that the constrained Cartan spinor obeys the eigenvalue equation $\Gamma'_{A}X^{A}\Psi' = \Psi'$, as in odd dimension. The symplectic structure of the $SU(2)$ doublet, implies the existence of two special invariant matrices: the identity and the symplectic form $J = \epsilon_{ij}$, which exchanges the two Pauli spinors. One is free to perform a similarity transformation $U$ to a more convenient basis. To make contact with the odd dimensional case, we reorder the components of $\varPsi$ using 
\begin{align}
    U = \frac{1}{\sqrt{2}}
    \begin{pmatrix} 1 && 0 && 0 && 0 \\ 
        0 && 0 && -1 && 0 \\
        0 && 1 && 0 && 0 \\
        0 && 0 && 0 && 1
    \end{pmatrix} \, .
\end{align}
We then have $\Gamma_A = U \Gamma'_{A}U^{\dagger}$, as in \crefrange{eq:Gammai}{eq:Gammaf}. We also obtain a supplementary matrix through  $U J U^{\dagger} \equiv -i\Gamma_{\star}$. Hence, although we are in odd ambient dimension, we have a coordinate independent construction of $\Gamma_{\star}$, the uplift of $\gamma_{\star}$ in the even dimensional slice. The output of this discussion is that the formalism defined for dS$_{2n+1}$ can be reused without modification for dS$_{2n}$, as long as one considers Dirac fields, while chiral fields may be considered using the more intricate chiral picture outlined at the start. 

  \subsection{Propagator and boundary limit}
    
From our kinematic discussion, we have a dimension independent formalism to uplift Dirac spinors to the ambient space. The most general 2-point function, compatible with the constraints on the polarisations and the homogeneity requirements, may only include the following structures
\begin{equation}
    \expval{\bm{\Psi}(X,\overline{S}_1)\bm{\overline{\Psi}}(Y,S_2)}= \overline{S}_1 S_2 g_+ (z) + \overline{S}_1 \Gamma_\star S_2 g_- (z)  \, ,
\end{equation}
We parameterise our functions as before in terms of the dS invariant variable, $z=\frac{1}{2}(1+X\cdot Y)$. We can also use the explicit expressions we found to project these structures down to a dS slice expression if required. For example, in planar coordinates we find
\begin{align}
     \overline{S}_1 S_2 &= \frac{\overline{s}_1 \gamma_\mu(x-y)^\mu s_2}{\sqrt{x^0 y^0}}  \label{eq:structures} \, , \\
     \overline{S}_1 \Gamma_\star S_2 &= \frac{\overline{s}_1 i \gamma_0 \gamma_\mu(\tilde{x}-y)^\mu s_2}{\sqrt{x^0 y^0}} \, ,
\end{align}
 where $\tilde{x}^{\mu} = (-x^0,x^a)$ is the time-reversal of $x^{\mu}$. In order to uplift the Dirac equation, in terms of a spin covariant derivative in the slice, we require a thorough mathematical treatment of the spin connection, such as that found in \cite{Trautman:1992,Trautman:1995fr}. We present an intuitive argument which reaches the same conclusion. The operator $\slashed{\nabla}$ should be a first order differential operator, longitudinal to the dS slice, whose image acting on a dS spinor is again, a dS spinor. The first requirement tells us that $\nabla_{A}$ includes only transverse objects, for example $G_{AB}\partial^{B}$ and $\Sigma_{AB}X^{B}$. The second requirement tells us that $\{\slashed{\nabla},\slashed{X}\}=0$. A minimal ansatz coming from the first constraint is easily fixed using the second one, to give us the final form of the Dirac operator in the embedding 
\begin{align}
    \slashed{\nabla} &= \Gamma^{A}\left(G_{AB}\partial^{A}-\Sigma_{AB}X^{B}\frac{1}{X\cdot X}\right) \\
    &= \slashed{\partial} - \slashed{X}X\cdot \partial - \frac{d+1}{2}\slashed{X} \, .
\end{align}
This offers a convenient derivation of the result stated in \cite{Nishida:2018opl,Trautman:1992}. One can explicitly check in the planar parameterisation that this operator reproduces the action of the covariant derivative on spinor as can be computed from the tetrad formalism \cite{Henningson:1998cd}
\begin{align}
    \Gamma^{A}\nabla_{A}\Psi &= \begin{pmatrix}
        \gamma_0 \\ - \slashed{x}
    \end{pmatrix}\frac{\gamma^{\mu}\nabla_{\mu}\psi}{\sqrt{-\eta}} \, , \\
    \gamma^{\mu}\nabla_{\mu}\psi&= \eta\gamma^{\mu}\partial_\mu \psi +\frac{d}{2}\gamma_{0}\psi \, .
\end{align}
It follows that $\bar{\Psi}\slashed{\nabla}\Psi = -2  \bar{\psi}\slashed{\nabla}\psi$. Similarly $\bar{\Psi}\Psi =0$, hence the mass term must be uplifted with a factor of $\Gamma_{\star}$, transforming $\Psi$ into a spinor with eigenvalue $\alpha = -1$. Explicitly, we find $\bar{\Psi}\Gamma_{\star}\Psi=2i \bar{\psi}\psi $. All together
\begin{align}
    \bar{\psi}\left(\slashed{\nabla}+m \right)\psi \equiv-\frac{1}{2}\bar{\Psi}\left(\slashed{\nabla}+im\Gamma_{\star}\right)\Psi\, ,
\end{align}
from which we can rewrite the Dirac equation in the ambient space as
\begin{align}
    \left(i\Gamma_{\star}\slashed{\nabla}-m\right)\Psi = 0 \, .
\end{align}
This form is used so that $\bar{S} \left(i\Gamma_{\star}\slashed{\nabla}-m\right)\Psi$ is of the same type as $\bar{S}\Psi$, and we may write $\bar{S}i\Gamma_{\star}\slashed{\nabla}\pdv{}{\bar{S}}-m$. For the case $m\neq 0$, in odd ambient space, the explicit factor of $\Gamma_{\star}$ prevents us from writing the Dirac equation of a single constrained Pauli spinor, as is clear from the representation-theoretic perspective.

One may also make use of the Casimir equation previously discussed. From the CFT literature, for example \cite{Isono:2017grm}, the Casimir of the $d$ dimensional Conformal group associated to fields transforming in a spin $\frac{1}{2}$ representation, is known and given by 
\begin{equation}
    C_{\Delta,\frac{1}{2}}= \Delta(d-\Delta)+\frac{d(d-1)}{8} \,  \footnote{This can be rederived by using the explicit form of the Casimir operator and considering the action of the Casimir on primary states in a $CFT_{d}$}.
\end{equation}
The generators of rotations for the spinor are given by $\Sigma_{AB}=\frac{1}{4}[\Gamma_{A},\Gamma_{B}]$, and the Casimir is built starting from the differential realisation of $L_{AB}$ on fields, 
\begin{equation*}
	 X_{A}\pdv{}{X^{B}}-X_{B}\pdv{}{X^{A}}+\overline{S}\Sigma_{AB}\pdv{}{\overline{S}} \, .
\end{equation*}
We can now evaluate $\mathcal{C}=-\frac{1}{2}\left(\mathcal{L}+ \Sigma\right)_{AB}\left(\mathcal{L}+ \Sigma\right)^{AB}$,

\begin{align}
    \mathcal{C}&= \underbrace{-\partial^2+X_A X_B\partial^A \partial^B+(d+1) X\cdot \partial}_{- \Box_{dS}}+ \underbrace{\frac{(d+2)(d+1)}{8}}_{-\frac{1}{2}\Sigma^2} + \underbrace{X\cdot \partial - \bar{S}\left(\Gamma^A X_A\right) \left(\Gamma^B \partial_B\right)\pdv{}{\bar{S}}}_{-\Sigma \cdot \mathcal{L}}\, .
\end{align}
Almost all of the pieces act trivially on the structures we defined in \cref{eq:structures}. In fact, only the last element of the third piece changes between the two structures appearing in the propagator, by picking up a sign. More importantly, they are completely decoupled. 

It is now a straightforward exercise to act with the Casimir equation on the propagator ansatz and to collect terms. The differential equation we obtain is simple, owing largely to the choice of variable 
\begin{equation}
    \left(z(z-1)\partial_z^2-\bigg(\frac{d+2-\alpha}{2}-(d+2)z\bigg)\partial_z-\left(\Delta+\frac{1}{2}\right)\left(d+\frac{1}{2}-\Delta\right)\right)g_\alpha(z)=0 \, .
\end{equation}
These hypergeometric differential equations are equivalent to the set of coupled, first order equations one finds using the Dirac equation 
\begin{align}
    (z-1)\partial_{z}g_{+}(z)+\frac{d+1}{2}g_{+}(z)+i m g_{-}(z)&=0 \, , \\
    z\partial_{z}g_{-}(z)+\frac{d+1}{2}g_{-}(z)+im g_{+}(z)&=0 \, ,
\end{align}
provided we identify $\Delta = \frac{d}{2}+i m$. The Dirac operator squaring to the Klein-Gordon equation tells us that Dirac spinors can only be in the principle series as we describe in \cref{app:repthy}. These differential equations are solved by hypergeometric functions, just as in the case of the scalar field in \cref{eq:scalarsol}. Solving the Dirac equation while requiring that all singularities lie at coincident point, fixes both functions in terms of one constant $\kappa_{\psi}$ : 
\begin{align}
\begin{split}
    g_{+}(z)&=\kappa_{\psi} \, _2F_1\left(d-\Delta +\frac{1}{2},\Delta +\frac{1}{2};\frac{d+1}{2};z\right) \, , 
    \end{split}\label{eq:spinW+}\\
    g_{-}(z)&= \kappa_{\psi} \frac{(d-2 \Delta )}{d+1} \, _2F_1\left(d-\Delta +\frac{1}{2},\Delta +\frac{1}{2};\frac{d+1}{2}+1;z\right) \, .\label{eq:spinW-}
\end{align}
As previously, the Hadamard condition fixes the leading singularity of the Wightman function to the same normalisation as in flat-space, where it diverges as  $-\frac{1}{4}\Gamma\left(\frac{d+1}{2} \right)\pi^{-\frac{d+1}{2}}u^{-\frac{d+1}{2}}\gamma_{\mu}\partial^{\mu}u$. One can check that all singularities contained in $g_{-}(z)$ are subleading, hence only $g_{+}(z)$ contributes, as we expect from the explicit coordinate form of the spinorial structures $\overline{S}_1 S_2$ and $\overline{S}_1 \Gamma_{\star}S_2$. Matching the overall constant, we find that the normalisation is given by 
\begin{align}
    \kappa_{\psi}= -\frac{1}{4}\frac{\Gamma \left(\Delta +\frac{1}{2}\right) \Gamma \left(d-\Delta +\frac{1}{2}\right)}{(4 \pi )^{\frac{d+1}{2}} \Gamma \left(\frac{d+1}{2}\right)} \, .
\end{align}
Note that, in the massless limit $\Delta \rightarrow \frac{d}{2}$, $g_{-}(z)\rightarrow 0$, while $g_{+}(z)$ is finite. From this expression for the Wightman function, one can reconstruct the different propagators to be used in perturbation theory as previously explained for the scalar case and given in \crefrange{def:ll}{def:lr}.

For the boundary limit of the 2-point function we proceed as in the scalar case, by parameterising the bulk in term of boundary vectors $X=\lambda P + \ldots$. We note that the spinor polarisations have a nontrivial power-law divergence as one approaches the boundary 
\begin{align}
    \lim_{\lambda \rightarrow \infty }\frac{1}{\sqrt{\lambda}}\overline{S}&=\overline{S}_{\partial} \, , \\
    \overline{S}_{\partial}\slashed{P}&=0  \, .
\end{align}
This can be checked explicitly for the case of conformally flat coordinates, and by homogeneity this must be true for any other limiting procedure toward the boundary. The boundary limit of the 2-point function at late time is given by 
\begin{align}
\begin{split}
    \lim_{\lambda \rightarrow \infty}\Pi(X=\lambda P +\ldots,Y, \overline{S}_1,S_2)&= \lambda^{-\Delta}\left(K_{\Delta}\frac{\overline{S}_{1,\partial}(1+\Gamma_{\star})S_2}{\left(-2P \cdot Y \right)^{\Delta+\frac{1}{2}}}+\mathcal{O}(\lambda) \right) \\
    &+\lambda^{\Delta-d}\left(K_{d-\Delta}\frac{\overline{S}_{1,\partial}(1-\Gamma_{\star})S_2}{\left(-2P \cdot Y \right)^{d-\Delta+\frac{1}{2}}}+\mathcal{O}(\lambda)\right) \, , 
\end{split}\label{eq:spinWblim}
\end{align}
with $K_{\Delta}\equiv -\frac{1}{8 \sqrt{\pi }^{d+2}}\Gamma \left(\Delta +\frac{1}{2}\right) \Gamma \left(\frac{d+1}{2}-\Delta \right)$. These again match the form expected for the bulk-to-boundary correlation function of spinors with weight $\Delta$ and $d-\Delta$.
    
\section{Discussion\label{sec:discussion}}

We conclude with a short description of the potential extensions and applications of the above formalism and results. 
\begin{itemize}
\item The most obvious extension of the techniques discussed above is the study of a greater range of fields of mixed integer and half integer spin, including those with gauge redundancy. In particular, it would be interesting to apply these techniques for fields describing the massless and partially massless UIRs described in \cite{Anninos:2017eib,Deser2001,Deser:2003gw,Baumann2018}, which are particular to dS. Another potential extension of our work would be to generalise to dS the study of weight shifting operators which simplify calculations in higher spin CFT and AdS \cite{Karateev:2018uk,Costa:2018mcg}. In \cite{Baumann:2019oyu} they touch on this subject, and extending this discussion to spinors and away from the late-time limit should prove worthwhile. Finally, while we worked with pure dS above, our explicit results should be useful to consider spinors in asymptotically dS spacetime, an important area for further analysis \cite{Sachs:1962zza,Arnowitt:1962hi,Anninos:2010zf,Anninos:2011jp}.

\item We also have a significant interest in the use of the ambient formalism to study the relation between QFT in fixed dS$_{d+1}$ and S$_{d+1}$ \cite{Visser:2017atf, Schlingemann:1998cw,Schlingemann:1999mk,David:2021wrw}. The time-like coordinate relates the respectively lorentzian and euclidean field theoretic properties of these spaces for both global and static coordinates described in \cite{anninos2012, Banks:2006rx}. In addition, there is a relation which encodes the thermodynamics of the cosmological horizon in the euclideanisation of the static patch \cite{Law:2020cpj,Muhlmann,Anninos2021, anninos2019Holo,Harris2021,Anninos:2020hfj}. The consequences of Wick rotation for propagators on these spaces should be relevant to this effort, as would a better understanding in general of the euclidean theory of $S_d$, perhaps building on  \cite{Schlingemann:1999mk}. 

\item The findings of this paper complement and further an expanding body of research on the development of a rigorous, analytic and group theoretic treatment of QFT in dS. The work of \cite{Sengor:2019mbz,Sengor:2021zlc} on the unitarity, and existence of scalar operators with a range of weights, and the comparison of their 2-point function to those of scalar fields and their conjugate momenta; is a pertinent example of the type of analysis to which it would be possible to apply our spinor construction. We have included statements on the limiting behaviour of the Wightman function in the late time regime for spinors in \cref{eq:spinWblim}, which may be directly compared with allegorical 2-point functions of spin half primaries constructed using these methods.

\item The structures found in planar coordinates for the spinor correlator are akin to the ones encountered in BCFT \cite{herzog2017}. Pushing this analogy further would be an interesting pursuit. One could also, in the Euclidean picture, consider the construction of isometry generators through topological surface operators, as in BCFT \cite{Herzog:2021spv}, with the hope of new insights on Ward identities and asymptotic symmetries in dS \cite{Sachs:1962zza,Arnowitt:1962hi,Anninos:2010zf,Anninos:2011jp}. Another possible extension is to make contact between our formalism and the one developed for massless fields \cite{David2019}, and the similar spinor helicities for CFT$_3$ \cite{Caron-Huot:2021kjy}.

\item We contextualised our work in the ongoing effort on the perturbative front, using the in-in formalism \cite{maldacena2003,Weinberg:1995mt,DiPietro:2021sjt,Sleight:2021plv,Gorbenko:2019rza}. The generalisation of these works to fermionic fields and tensors is a natural objective, for example, in the construction of the effective AdS action studied in detail for scalar fields in \cite{DiPietro:2021sjt}. Simplification of perturbative calculations in dS is achieved there by constructing a non-unitary Lagrangian in an AdS background which reproduces the dS results at each order in perturbation theory. Additionally, the cosmological bootstrap effort offers hope for another application of the ideas included in our work \cite{Sleight:2019hfp,Sleight:2019mgd,McFadden2010b,Hogervorst:2021uvp,Baumann:2020ksv,Sengor:2019mbz}, perhaps enlarging our knowledge of CFT like structures in dS beyond the previous work on the dS/CFT correspondence \cite{Anninos:2011ui,Witten:2001kn, Strominger:2001pn,Harlow:2011ke}. In particular, the development of the K\"all\'en-Lehman spectral representation and the expansion of the four-point function of boundary operators in terms of conformal partial waves will have relevance for the study of spinor theories in analogy to the scalar case.  
\end{itemize}

\acknowledgments

We would like to thank D.~Anninos and C.~Herzog for their support and encouragement throughout this project. Their comments and suggestions for improvement on the various drafts greatly helped us, and they gave us the confidence to bring this project to completion. We would like to thank K.~Nguyen, P.~Benetti Genolini, T.~Bautista, E.~Harris, M.~Downing, S.~Sheorey  and T.~Orchard for the stimulating and enriching discussions, as well as J.~Phillips for offering us his insight on differential geometry and spinors. BP would like to thank the STFC for support under grant ST/W507556/1 .

\appendix
\section{Conventions}\label{app:Conv}

In all of this paper, the indices considered are flat, i.e. they are contracted using Minkowski or euclidean metric, depending on the range. Indices are never contracted using the curved-space metric $g_{\mu\nu}$.  We use the mostly-plus convention. Latin indices from the start of the alphabet designate the ($d$)-spatial coordinate, $a,b,c, \ldots = 1, 2, \ldots d$. Latin indices from the middle of the alphabet range from $1$ to $d+1$, useful when considering global coordinates where angular variables  $\omega_{i}\omega^{i}=1$ appear. Greek indices are used in the usual fashion, $\mu,\nu ,\rho = 0,1, \ldots d$. We use upper-case latin indices for the embedding coordinates in $\mathbb{R}^{1,d}$, i.e. $A,B,C \ldots = 0,1, \ldots d+1$. Equivalently, we make use of light-cone variables and metric, where $A,B,C \ldots = +,-, 1, \ldots d$. In our parametrisation, $\eta_{+,-}=-\frac{1}{2}$, i.e. $X^{\pm}=X^{0}\pm X^{d+1}$. Antisymmetrisation and symmetrisation of indices are written using respectively square and round bracket, and have weight $1$, i.e. $T_{(ab)}= \frac{1}{2}\left(T_{ab}+T_{ba}\right)$.

The action of symmetry generators $\hat{P},\hat{K},\ldots$ on fields $\Phi$ are realised through some differential operators $\hat{Q}_{P},\hat{Q}_{K}, \ldots$ defined as  
\begin{align}
    [\hat{P},\Phi] = \hat{Q}_{P}\cdot \Phi\, .
\end{align}

It follows from the Jacobi-identity that the $\hat{Q}$ have commutation relations given by minus those of the operators they represent, i.e. $\hat{Q}_{[P,K]}=-[\hat{Q}_{P},\hat{Q}_{K}]$. These considerations are important should one wish to reproduce the detail of the derivations of the spin part of generators acting on fields on the slice. The generators of rotations in the embedding space are characterised by commutation relations and realisation
\begin{align}
    [L_{AB},L^{CD}]&=-\eta_{A}{}^{C}L_{B}{}^{D}+\eta_{B}{}^{C}L_{A}{}^{D}+\eta_{A}{}^{D}L_{B}{}^{C}-\eta_{B}{}^{D}L_{A}{}^{C}  \\
    &=-4 \eta_{[A}{}^{[C}L_{B]}{}^{D]} \, , \\
	[L_{AB}, \Psi(P) ] &=- (\mathcal{L}_{AB}+\Sigma_{AB})\Psi(P) \, , \\
	\mathcal{L}_{AB} &= P_A \pdv{}{P^{B}}-P_B \pdv{}{P^{A}} \, .
\end{align}
$L_{AB}, \mathcal{L}_{AB}$ and $\Sigma_{AB}$ all obey the same commutation relations. $\Sigma_{AB}$ are the usual spin-matrices that give matrix representation of $SO(1,d+1)$ or $Spin(1,d+1)$. The operators $L_{AB}$ are anti-hermitian. This algebra can be repackaged in multiple fashions, two of which are useful for our purpose. The (euclidean) conformal algebra is identified through
\begin{align}
	D &= 2L_{+,-}\, ,  \\
	P_a &= 2L_{-,a}\, ,  \\
	K_a &= 2L_{a,+}\, ,  \\
	M_{ab} &= L_{ab} \, , 
\end{align}
and its commutation relations follow straightforwardly,
\begin{align}
	[D,P_{a}] &= P_{a}\, ,  \\
	[D,K_{a}] &= -K_{a}\, ,  \\
	[K_{a},P_{b}] &= 2 \delta_{ab}D-2M_{ab}\, ,  \\
	[M_{ab},P^{c}] &= -2\delta_{[a}^{c}P_{b]}\, ,  \\
	[M_{ab},K^{c}] &= -2\delta_{[a}{}^{c}K_{b]}\, ,  \\
	[M_{ab},M^{cd}]&= -4\delta_{[a}{}^{[c}M_{b]}{}^{d]}\, , 
\end{align}
with all other commutators vanishing. The quadratic Casimir is given by $\mathcal{C}=-\frac{1}{2}L_{AB}L^{AB}= D^2 + \frac{1}{2}\left(P\cdot K+ K \cdot P \right)- \frac{1}{2}M_{ab}M^{ab}$. Its eigenvalue can be found by considering primary fields of a $CFT_{d}$. The scaling part gives the usual $\Delta(d-\Delta)$, while the spin part for tensors of spin-$j$ gives $j(j-d+2)$, and for Dirac spinors $\frac{d(d-1)}{8}$. 

The de Sitter algebra is identified by separating rotations and boosts 
\begin{align}
    K_{i}&=L_{0i} \, , \\
    M_{ij}&=L_{ij} \, .
\end{align}
The commutation relations of the algebra follow straightforwardly 
\begin{align}
    [M_{ij},M^{kl}]&= -4 \delta_{[i}{}^{[k}M_{j]}{}^{l]}  \, ,\\
    [M_{ij},K^{k}]&=-2 \delta_{[i}{}^{k}K_{j]} \, , \\
    [K_i,K_j] &= M_{ij} \, .
\end{align}

We find the language of \cite{Trautman:1995fr} convenient to refer to the different spinor representations. In even dimensions, the fundamental spinors are the left-handed (LH, $+$) and right-handed (RH,$-$) Weyl spinors. A Dirac spinor is the direct sum of a left and right Weyl spinor. In odd dimension, the irreducible representation (which is of Dirac type), is called a Pauli spinor. An $SU(2)$ doublet of Pauli spinors, one in each inequivalent representation of the Clifford algebra, form a Cartan spinor. This is of course a reducible representation, and the odd-dimensional analogue of the Dirac representation. We prove the following relations: Dirac spinors in dS$_{2n}$ are uplifted to constrained Cartan spinors of $Spin(1,2n)$, while Pauli spinors in dS$_{2n+1}$ are uplifted to constrained Dirac spinors of $Spin(1,2n+1)$. In all cases, we consider a set of gamma matrices obeying $\{\gamma_{\mu},\gamma_{\nu}\}=2\eta_{\mu\nu}$. The spin-matrix is then given by $\Sigma_{\mu\nu}=\frac{1}{4}[\gamma_{\mu},\gamma_{\nu}]$. In the embedding space, we use $\Gamma_{A}$ instead. Conjugation properties follow from $\gamma_{\mu}^{\dagger} = \gamma_0 \gamma_{\mu}\gamma_{0}$. Conjugate spinors are defined through $\overline{\psi}=\psi^{\dagger}i\gamma_0$, and similarly in the embedding. This choice matches that of Weinberg \cite{Weinberg:1995mt}, such that $\left(\overline{\alpha}\beta\right)^{\star} = \overline{\beta}\alpha$ and $\left(\overline{\alpha}\gamma_\mu\beta\right)^{\star} = -\overline{\beta}\gamma_\mu\alpha$. In even dimensions, we write the chiral matrix with $\gamma_{\star}^2=1$, and we use chiral-$\gamma$ matrices, or $\sigma$-matrices. This means we consider 
\begin{align}
     \gamma_\mu &=\begin{pmatrix} 0 && \sigma_\mu \\ \overline{\sigma}_\mu && 0 \end{pmatrix} \, , \\ 
    \sigma_\mu &= (\mathds{1}, \sigma_a) = \sigma_\mu^\dagger \, , \\
    \overline{\sigma}_\mu &= (-\mathds{1}, \sigma_a) = \overline{\sigma}_\mu^\dagger \, , \\
    \sigma_{(\mu}\overline{\sigma}_{\nu)}&=\eta_{\mu\nu} \, , \\
    \gamma_{\star} &=\begin{pmatrix} \mathds{1} && 0 \\ 0 && -\mathds{1} \end{pmatrix} \, , \\
    \mathbb{P}_{\pm}&=\frac{\mathds{1}\pm \gamma_\star}{2} \, .
\end{align}
The $\sigma$ matrices also give rise to chiral-rotation matrix, $\sigma_{\mu\nu}=\frac{1}{2}\sigma_{[\mu}\overline{\sigma}_{\nu]}$ and $\overline{\sigma}_{\mu\nu}=\frac{1}{2}\overline{\sigma}_{[\mu}\sigma_{\nu]}$, which appear when considering Weyl spinors. Embedding spinors are generically named $\Psi$ and written in block form 
\begin{align}
	\Psi &= \begin{pmatrix}
		\chi \\
		\rho \end{pmatrix} \, , \\
	\overline{\Psi}&= \begin{pmatrix}
		\overline{\rho}\gamma_0 &&
		-\overline{\chi}\gamma_0 \end{pmatrix} \, .
\end{align}
When considering their transformation law, one has to take into account a sign and ordering difference
\begin{align}
	[L_{AB}, \Psi(P) ] &=-\mathcal{L}_{AB}\Psi(P)-\Sigma_{AB}\Psi(P) \, , \\
	[L_{AB}, \overline{\Psi}(P) ] &=-\mathcal{L}_{AB} \overline{\Psi}(P)+ \overline{\Psi}(P)\Sigma_{AB}\, ,
\end{align}
and similarly for the commutation relations on the dS slice. dS spinors are usually named $\psi$, and are related non-trivially to $\chi$ and $\rho$, as shown in the text. Chiral spinors are preferably encoded using a whole Dirac spinor with eigenvalue equation $\gamma_{\star}\psi_{\pm} = \pm \psi_{\pm}$.

\section{Representation Theory \label{app:repthy}} 

In this appendix we collect some results regarding the representation theory of the de Sitter group, $SO(1,d+1)$. For a more thorough treatment, one should refer to the canonical work \cite{Dobrev:1977qv}, or to the excellent recent review of the subject \cite{Sun2021}. Spinorial representations are only lightly touched in the literature, but we present some interesting elementary statements regarding them \cite{Thieleker1973,Thieleker1974}.

The representation theory of de Sitter has some similarities to the familiar examples of conformal field theory and (E)AdS. The operators we are concerned with transform in finite dimensional representations of the de Sitter group, noting that these cannot be unitary representations due to their finite dimensionality. In this paper we consider Symmetric Traceless Tensors of generic integer spin $J$, and spin $J=\frac{1}{2}$ representations of the universal cover of the ambient space Lorentz group, which acts as the de Sitter group once pulled to the de Sitter slice. The action of this group is described in the ambient space formalism in \cite{Sun:2020sgn}. The spectrum of single particle states of the fields should transform in unitary irreducible representations of the de Sitter group. To induce the irreducible unitary representations we consider the maximal subgroup $SO(1,1) \times SO(d)\in SO(1,d+1)$, and label representation using label $(\Delta,\rho)$ with $\rho$ a representation of $\mathfrak{so}(d)$. Unitarity imposes a complex interplay between the spin representation and the allowed values of $\Delta$. These organise into two continuous series and two discrete \cite{Sengor:2019mbz,Deser:2003gw,Joung2006,Joung2007,Anous2020,Newton1950}. For clarity we treat the symmetric traceless tensors and the Dirac spinor cases separately.

\subsection{Symmetric traceless tensors}

Irreducible representations of $\mathfrak{so}(d)$ are specified by a weight-vector $\vec{s} = (s_1,s_2,....s_r)$, of dimension $r=\lfloor\frac{d}{2}\rfloor$, with half integer entries $s_i \in \frac{1}{2} \mathbb{N}$ \cite{Basile2017}. For the bosonic case  $\vec{s}$ defines a young tableaux with rows of length $s_1\geq s_2\geq ....\geq s_r$. Symmetric Traceless Tensor (STT) representations correspond to young tableau with one row and $J \in \mathbb{N}$ boxes, i.e. $\vec{s} = (J,0,...,0)$. Let $p$ be the number of non zero entries in $\vec{s}$, so that only the spin 0 case is distinguished. The induced unitary irreducible representations (UIRs) are entirely specified given the spin representation and the value of $\Delta$. We additionally identify representations with weight $\Delta$ to those with weight $d-\Delta$. These  are unitary in de Sitter as well as equivalent to those of weight $\Delta$, under the intertwining isomorphism given by the `shadow transformation', which is discussed in \cite{Sun2021,Sengor:2019mbz,Simmons-Duffin:2014wb}. The allowed value of $\Delta$ for a given STT representation then decomposes into the following series :
\begin{itemize}
    \item Principal series : $\Delta = \frac{d}{2}+i\nu$, $\nu \in \mathbb{R}^{+}$
    \item Complementary series : $\Delta \in(\frac{d}{2},d-p)$ 
     \item Exceptional series : $(\Delta +J-2)(d+J-2-\Delta) = (J-1-t)(d+J+t-3)$ for $t \in 0, 1,... ,J-1$.
     \item Discrete series : for $d+1 = 2n$, $\Delta = \frac{d}{2} + \frac{1}{2}\mathbb{N}$
\end{itemize}
The conformal dimension $\Delta$ can be related to the mass of the fields and states in de Sitter. The most natural choice of the mass parameter for various representations in de Sitter is the subject of \cite{Deser:2003gw}, and leads to a description of unique representations in the case of de Sitter for higher spin fields. In particular there exist `massless' and `partially massless' representations, associated with the exceptional series. These are discussed at length in \cite{Anninos:2017eib,Baumann2018,Deser2001,Basile2017,Higuchi:2010xt,Joung2016}. The latter are unitary, in the above sense, and correspond to discrete values of the mass between the lower bound of the complementary series and $m=0$. These partially massless representations have an intermediate amount of gauge freedom, interpolating the massless and massive cases. 
For STT operators, the Laplacian is equal to the Casimir up to a constant shift. For this reason in \cref{app:STT} we solve for the Wightman function using the equation of motion according to \cite{Costa:2014kfa}
\begin{equation}
    (\Box_{dS}-\Delta(d-\Delta) -J )\Pi = 0 \, .
\end{equation}

\subsection{Spinors}
Local spinor fields transform in the familiar spin $\frac{1}{2}$ representations of the local Lorentz group pulled back from the ambient space \cite{VanProeyen:1999ni,Weinberg:1995mt}. They may be constructed by demanding they solve the Dirac equation with the appropriate spin covariant derivative \cite{Deser:2003gw,Cotaescu2002,Cotaescu2018,Stahl2016}. The Dirac equation imposes a specific form of the mass term in the Laplacian eigenproblem \cite{Candelas1975,Deser2001,Isono:2017grm}
 \begin{equation}
     (\slashed{\nabla} - m)\psi = 0 \implies \left(\Box_{dS} -\left(m^2 + \frac{R}{2}\right)\right)\psi = 0 \, .
 \end{equation}
From which it follows that, in terms of the UIRs, we can only find
\begin{itemize}
    \item Principal series : $\Delta = \frac{d}{2}- i m$
    \item Exceptional series : $\Delta = \frac{d}{2}$.
\end{itemize}
Complementary series single particle UIRs of the de Sitter group are excluded from the spectra of half integer spin representations of the fields \cite{Thieleker1973,Thieleker1974}. This nontrivial point relies on the observation that the fermionic UIRs, induced by those of the double cover of the compact subgroup, $Spin(d)$, are incompatible with the positivity of the intertwining operator, while it is required to write a unitary inner product for the complementary series. This follows from the general statement made in \cite{Thieleker1974}, that positivity requires the inducing representation of $Spin(d)$ to be equivalent to its Weyl conjugate. In the case $d = 2p$, this excludes the half integer spin representations of $Spin(d)$, which are necessarily chiral. In the case, $d = 2p+1$, faithful (injective) representations of $Spin(d)$ are also excluded from the complementary series, by the requirement that the highest weight state must have a final entry with integer values. The injective representations of $Spin(d)$ are precisely those of half integer spin, as the bosonic representations are double valued. This exclusion can also be seen more mundanely as above. The shift in the mass term when squaring the Dirac equation, changes the relation between $\Delta$ and $m$ encountered for bosons. The complementary series then requires an imaginary value of $m$, which implies non-unitarity of the Hamiltonian. The works previously cited  include an analysis of higher rank spinor tensors and introduce the possibility of partially massless tunings for $m$ for these fields, we leave this topic for later analysis. 
 
\section{STT Wightman Function} \label{app:STT}

The STT propagator in AdS has been calculated in \cite{Costa:2014kfa}. Here we present a calculation of the Wightman function for symmetric traceless tensor fields in analogy. The case of symmetric traceless tensors in (A)dS, as well as the link to the  euclidean sphere $S^N$ and hyperbola $H^N$ has been treated in \cite{Sun:2020sgn, Sleight:2017fpc,Higuchi:1985ad, Higuchi1987,Higuchi1987cf} among others.  Here we perform the calculation explicitly in dS for reference and convenience, although the result follows in principle from the analytic continuation of the AdS Harmonic function\cite{Hogervorst:2021uvp}. The Wightman function between ambient space points $X_1$ and $X_2$ of a massive spin $J$ field in dS with polarisation vectors $W_1$ and $W_2$ respectively is dependent on the chordal distance $z = \frac{1}{2}(1+X_1\cdot X_2) = 1-\frac{u}{4}$. We make use of two equivalent bases
\begin{equation*}
\begin{aligned}
	\Pi_{J,\Delta}(X_1,W_1; X_2,W_2)&= \sum_{k=0}^J(W_1 \cdot W_2)^{J-k} ((W_1\cdot X_2)( W_2  \cdot X_1))^k g_k(z) \\
	&=  \sum_{k=0}^J(W_1 \cdot W_2)^{J-k} ((W_1\cdot \nabla_1) (W_2  \cdot \nabla_2))^k f_k(z).
\end{aligned}
\end{equation*}
\ni We can recover the first basis from the second via
\begin{equation}
    g_k(z)  =  \sum_{i=k}^J \left(\frac{1}{2}\right)^{i+k}\left(\frac{i!}{k!}\right)^2\frac{1}{(i-k)!} \partial_z^{(i+k)}f_i(z) \, .
\end{equation}
The equation of motion provides the necessary differential equation which we seek to solve, 
\begin{equation}
    (\Box - (\Delta(d-\Delta) + J )) \Pi_{J,\Delta}(X_1,X_2) = 0 \, .
\end{equation}
With the covariant derivative given as in \cref{eq:CovDiv}. This equation is equivalent to the Casimir equation. To further simplify, we define $h_k(z) = \partial_z^k f_k(z)$. The equation of motion can be written recursively :
\begin{align}
	\begin{split}
\Big( (1-z) z \partial_z^2+ (d+1+ 2k)\left(\frac{1}{2}-z\right)
  \partial_z- \hspace{2cm}& \\ \Delta  (d-\Delta )-2 k(k-J+1) \Big) h_k(z)&=4 (J-k+1) h_{k-1}(z) \, , \\
   h_{-1}(z)&= 0 \, .
	\end{split}
\end{align}
The $k=0$ equation is simply the equation of motion for the scalar field and so is easily solved and normalised as described in \cref{sec:scalar}. The following equations can be solved recursively from the two previous ones as in EAdS,
\begin{align}
\begin{split}\label{eq:recrusiv}
    h_k(z)&=c_{J,k}\left( (d+2 J-2 k-1) \left((2-d-J) h_{k-1}(z)+\left(\frac{1}{2}-z\right) h_{k-1}'(z)\right) \right.\\&\left.\hspace{8cm}+2 (J-k+2) h_{k-2}(z)\vphantom{\frac{a}{b}}\right),
\end{split}
\end{align}
Where
\begin{align*}
    c_{J,k}=\frac{-2(J-k+1)}{k (d+2 J-k-2) (\Delta +J-k-1) (d-\Delta +J-k-1)}\, .
\end{align*}
The recursion relations in AdS follow by changing variable $z\rightarrow -\sigma$ and the sign of the term multiplying $h_{h-1}$ in \cref{eq:recrusiv}.

\section{Spinors in Global Coordinates} \label{app:Global}

In the main text we showed that a generic Dirac-type spinor $\Psi$ of $\mathbb{R}^{1,d+1}$ with eigenvalue equation $\Gamma^{A}X_{A}\Psi = \Psi$ encodes a Dirac spinor of the dS slice. Our proof uses flat-slicing, planar coordinates, but holds in general by the homogeneity of dS$_{d+1}$. It is convenient to have coordinate expression for some practical computations, and the one used previously does not cover the whole space, nor do they make the analytic continuation to $S_{d+1}$ clear. This is why we devote this appendix to the analogous construction for global coordinates
\begin{equation}
    X^{A}=(X^{0},X^{i})=(\sinh(t),w^{i}\cosh(t)) \, ,
\end{equation}
with angular variables $w^{i}w_{i}=1$. These have the benefit of both covering the whole space and making the analytic continuation $t=i\tau$  straightforward. These coordinates treat the $d+1$-th component indistinctly from the others, and so the previous splitting of the $SO(1,d+1)$ Lorentz algebra into the conformal algebra is ill-suited to analyse the transformation law of fields induced on the slice. It is more natural to use a different decomposition of the group by identifying the de Sitter algebra
\begin{align}
    K_{i}&=L_{0i} \, , \\
    M_{ij}&=L_{ij} \, .
\end{align}
Note that this $K_{i}$ is unrelated to the special conformal transformation generator of the conformal algebra. This splitting simply isolates the boosts and the rotations, and is precisely the one used to study the representations of $SO(1,3)$ in \cite{Weinberg:1995mt}. The commutation relations follow directly
\begin{align}
    [M_{ij},M^{kl}]&= -4 \delta_{[i}{}^{[k}M_{j]}{}^{l]} \, , \\
    [M_{ij},K^{k}]&=-2 \delta_{[i}{}^{k}K_{j]}  \, , \\
    [K_i,K_j] &= M_{ij} \, .
\end{align}

From the explicit form of the coordinate slices, we can find that acting on a scalar field $\phi(X(t,w))$ in the embedding, they  act as the Killing vectors 
\begin{align}
    \hat{Q}_{M_{ij}}\phi(X(t,w))&= w_{j}\pdv{}{w^{i}}\phi-w_{i}\pdv{}{w^{j}}\phi \, , \\
    \hat{Q}_{K_{i}}\phi(X(t,w))&= w_i \pdv{}{t}\phi+\tanh(t)\underbrace{\left(\delta_{ij}-w_{i}w_{j}\right)\pdv{}{w_{j}}}_{=h_{ij}\partial^{j}=\nabla_{i}}\phi \, .
\end{align}
We see that the $M_{ij}$ implement rotations while the boosts, $K_{i}$, contain both a time translation as well as a covariant derivative on the sphere. The covariant derivative is unsurprising as it is necessary for the generators to preserve the constraint $w^2=1$. Acting on a field in the slice $\psi(t,w)$ with definite spin representation specified by spin matrices $\Sigma_{AB}$, the orbital part of the generator for the scalar has to be supplemented by the spin part 
\begin{align}
    [M_{ij},\psi(t,w)]&= -\left(w_{i}\pdv{}{w^{j}}-w_{j}\pdv{}{w^{i}}+\Sigma_{ij}\right)\psi  \, , \\
    [K_{i},\psi(t,w)]&= w_i \pdv{}{t}\psi+\tanh(t)\nabla_{i}\psi+S_{i}(t,w,\Sigma)\psi \, .
\end{align}
The vector $S_{i}(t,w,\Sigma)$ is non-trivial to compute, although it is entirely fixed by the Jacobi identity $[[A,B],\psi]=[A,[B,\psi]]-[B,[A,\psi]]$. We proceed in two steps. First, note that the Jacobi identity for $[M,K]$ is solved by the general ansatz $S_{a}=\Sigma_{ij}w^{j}g_1(t)+\Sigma_{i0}g_2(t)$. Secondly, one can use this ansatz in the Jacobi identity for $[K,K]$ and decompose the resulting equation in terms of independent structures multiplied by equations involving $g_1(t)$ and $g_2(t)$, which must all vanish. The output of this is that the action of the symmetry generators on fields with spin do generate a representation of the de Sitter algebra provided $g_1(t)=-\frac{1}{\tanh(t)}$ and $g_2(t)=-\frac{1}{\sinh(t)}$. From this analysis, we gather that on the slice, a generic spinning field transforms under the action of the generator according to

\begin{align}
    [M_{ij},\psi(t,w)]&= -\left(w_{i}\pdv{}{w^{j}}-w_{j}\pdv{}{w^{i}}+\Sigma_{ij}\right)\psi  \, , \\
    [K_{i},\psi(t,w)]&=\left(w_i \pdv{}{t}+\tanh(t)\nabla_{i}-\frac{\cosh(t)\Sigma_{ij}\omega^{j}+\Sigma_{i0}}{\sinh(t)}\right)\psi \, .
\end{align}

This can, as previously, be compared with the transformation induced from the embedding on a generic spinor field $\Psi$ as in \cref{eq:SpinorRep}. As argued in the main text, one can consider the case $d+1=2k+1$, and use the resulting formalism in any dimension, both odd and even. A parameterisation of the gamma matrices can be chosen similarly as before
\begin{align}
    \Gamma_{0}&= \begin{pmatrix}
        0 && 1 \\ -1 && 0 
    \end{pmatrix}\otimes \mathds{1} \, , \\
     \Gamma_{i}&= \begin{pmatrix}
        1 && 0 \\ 0 && -1
    \end{pmatrix} \otimes\gamma_{0}\gamma_{i} \, .
\end{align}
Using this convention, one can check that rotations behave as expected, however boosts require more care. Motivated by the planar case, we consider the commutation relation not for $\chi$ itself but for $f(t)\chi$, and leave $f(t)$ to be determined : 
\begin{align}
    [K_{i},f(t)\chi]&=\Big(w_{i}\partial_t+\tanh(t)\nabla_i \Big)(f\chi)-w_i\dot{f}(t)\chi+\frac{1}{2}\gamma_0 \gamma_i f \rho \\
    &=\Big(w_{i}\partial_t+\tanh(t)\nabla_i \Big)(f\chi)-\frac{\gamma_i \slashed{w}-w_{i}}{2\tanh(t)}f(t)\chi-\frac{\gamma_i \gamma_0}{2\sinh(t)} f \chi \, .
\end{align}
Where the first line is derived from the explicit transformation law of the embedding space spinor, the second line is required for $f(t)\chi$ to transform as a dS$_{d+1}$ spinor. This equality is solved provided
\begin{align}
    2\tanh(t)\dot{f}(t)+f(t)&=0 \Rightarrow f(t)=\frac{1}{\sqrt{\sinh(t)}} \, , \\
  \rho &= \frac{1-\cosh(t)\gamma_0\slashed{w}}{\sinh(t)}\chi \, .
\end{align}
This, as expected, corresponds to an embedding spinor satisfying $\Gamma_{A}X^{A}\Psi = \Psi$. Proceeding as before, one can define a polarisation vector such that 
\begin{align}
    \overline{S}\Psi = \frac{1}{\sqrt{\sinh(t)}}\begin{pmatrix} \overline{s} && 0 \end{pmatrix}\left(\Gamma_{A}X^{A}+1\right)\Psi = \overline{s}\psi \, .
\end{align}
The kinematics for global coordinates is then entirely fixed by using polarisations 
\begin{align}
    X^A \Gamma_A S = -S  \hspace{1cm}&\implies\hspace{1cm}  S= \begin{pmatrix} -\gamma_0\sinh(t) \\ \gamma_0+\cosh(t)\slashed{w} \end{pmatrix}\frac{s}{\sqrt{\sinh(t)}} \, , \\
     \overline{S}X^A \Gamma_A = \overline{S}\hspace{1cm}&\implies\hspace{1cm} \overline{S}=\frac{\overline{s}}{\sqrt{\sinh(t)}}\begin{pmatrix} \cosh(t)\gamma_0 \slashed{w}+1 && \sinh(t) \end{pmatrix}  \, .
\end{align}
One can use these expressions to find the different spinorial structures appearing in the correlation function. It is also straightforward to check that the boundary limit of these polarisations is as stated previously, and once rescaled we obtain a smooth limit to the polarisation spinor of a primary spinor on the lightcone.

%----------------------------------------------------
%----------------------------------------------------
%----------------------------------------------------

%\nocite{*}

% Dio style bib :

%\section*{References}
%\printbibliography[keyword=D,title={De Sitter Space}]
%\printbibliography[keyword=A,title={Anti de Sitter Space}]
%\printbibliography[keyword=C,title={Conformal Field Theory}]
%\printbibliography[keyword=R,title={Representation theory and other mathematical background}]

% JHEP style bib 
\bibliographystyle{jhep}
\bibliography{paper.bib}

\providecommand{\href}[2]{#2}\begingroup\raggedright\begin{thebibliography}{100}

\bibitem{Spradlin:2001pw}
M.~Spradlin, A.~Strominger and A.~Volovich, \emph{{Les Houches lectures on de
  Sitter space}},  in \emph{{Les Houches Summer School: Session 76: Euro Summer
  School on Unity of Fundamental Physics: Gravity, Gauge Theory and Strings}},
  pp.~423--453, 10, 2001
  [\href{https://arxiv.org/abs/hep-th/0110007}{{\ttfamily hep-th/0110007}}].

\bibitem{Anninos:2012qw}
D.~Anninos, \emph{{De Sitter Musings}},
  \href{https://doi.org/10.1142/S0217751X1230013X}{\emph{Int. J. Mod. Phys. A}
  {\bfseries 27} (2012) 1230013}
  [\href{https://arxiv.org/abs/1205.3855}{{\ttfamily 1205.3855}}].

\bibitem{Baumann:2009ds}
D.~Baumann, \emph{{Inflation}},  in \emph{{Theoretical Advanced Study Institute
  in Elementary Particle Physics}: {Physics of the Large and the Small}},
  pp.~523--686, 2011, \href{https://doi.org/10.1142/9789814327183_0010}{DOI}
  [\href{https://arxiv.org/abs/0907.5424}{{\ttfamily 0907.5424}}].

\bibitem{Dirac:1935zz}
P.A.M.~Dirac, \emph{{The Electron Wave Equation in De-Sitter Space}},
  \href{https://doi.org/10.2307/1968649}{\emph{Annals Math.} {\bfseries 36}
  (1935) 657}.

\bibitem{Dirac:1936fq}
P.A.M.~Dirac, \emph{{Wave equations in conformal space}},
  \href{https://doi.org/10.2307/1968455}{\emph{Annals Math.} {\bfseries 37}
  (1936) 429}.

\bibitem{Weinberg:2010ws}
S.~Weinberg, \emph{Six-dimensional {Methods} for {Four}-dimensional {Conformal}
  {Field} {Theories}},
  \href{https://doi.org/10.1103/PhysRevD.82.045031}{\emph{Physical Review D}
  {\bfseries 82} (2010) 045031}.

\bibitem{Costa:2011wa}
M.S.~Costa, J.~Penedones, D.~Poland and S.~Rychkov, \emph{Spinning {Conformal}
  {Correlators}}, \href{https://doi.org/10.1007/JHEP11(2011)071}{\emph{Journal
  of High Energy Physics} {\bfseries 2011} (2011) 71}.

\bibitem{Costa:2011vf}
M.S.~Costa, J.~Penedones, D.~Poland and S.~Rychkov, \emph{Spinning {Conformal}
  {Blocks}}, \href{https://doi.org/10.1007/JHEP11(2011)154}{\emph{Journal of
  High Energy Physics} {\bfseries 2011} (2011) 154}.

\bibitem{Costa:2014kfa}
M.S.~Costa, V.~Gon\c{c}alves and J.a.~Penedones, \emph{{Spinning AdS
  Propagators}}, \href{https://doi.org/10.1007/JHEP09(2014)064}{\emph{JHEP}
  {\bfseries 09} (2014) 064} [\href{https://arxiv.org/abs/1404.5625}{{\ttfamily
  1404.5625}}].

\bibitem{Penedones:2010ue}
J.~Penedones, \emph{{Writing CFT correlation functions as AdS scattering
  amplitudes}}, \href{https://doi.org/10.1007/JHEP03(2011)025}{\emph{JHEP}
  {\bfseries 03} (2011) 025} [\href{https://arxiv.org/abs/1011.1485}{{\ttfamily
  1011.1485}}].

\bibitem{Penedones:2016voo}
J.~Penedones, \emph{{TASI lectures on AdS/CFT}},  in \emph{{Theoretical
  Advanced Study Institute in Elementary Particle Physics}: {New Frontiers in
  Fields and Strings}}, 8, 2016,
  \href{https://doi.org/10.1142/9789813149441_0002}{DOI}
  [\href{https://arxiv.org/abs/1608.04948}{{\ttfamily 1608.04948}}].

\bibitem{Costa:2018mcg}
M.S.~Costa and T.~Hansen, \emph{{AdS Weight Shifting Operators}},
  \href{https://doi.org/10.1007/JHEP09(2018)040}{\emph{JHEP} {\bfseries 09}
  (2018) 040} [\href{https://arxiv.org/abs/1805.01492}{{\ttfamily
  1805.01492}}].

\bibitem{Meltzer:2019nbs}
D.~Meltzer, E.~Perlmutter and A.~Sivaramakrishnan, \emph{{Unitarity Methods in
  AdS/CFT}}, \href{https://doi.org/10.1007/JHEP03(2020)061}{\emph{JHEP}
  {\bfseries 03} (2020) 061}
  [\href{https://arxiv.org/abs/1912.09521}{{\ttfamily 1912.09521}}].

\bibitem{Nishida:2018opl}
M.~Nishida and K.~Tamaoka, \emph{{Fermions in Geodesic Witten Diagrams}},
  \href{https://doi.org/10.1007/JHEP07(2018)149}{\emph{JHEP} {\bfseries 07}
  (2018) 149} [\href{https://arxiv.org/abs/1805.00217}{{\ttfamily
  1805.00217}}].

\bibitem{Sengor:2019mbz}
G.~Seng\"or and C.~Skordis, \emph{{Unitarity at the Late time Boundary of de
  Sitter}}, \href{https://doi.org/10.1007/JHEP06(2020)041}{\emph{JHEP}
  {\bfseries 06} (2020) 041}
  [\href{https://arxiv.org/abs/1912.09885}{{\ttfamily 1912.09885}}].

\bibitem{Sengor:2021zlc}
G.~Sengor and C.~Skordis, \emph{{Scalar two-point functions at the late-time
  boundary of de Sitter}},  \href{https://arxiv.org/abs/2110.01635}{{\ttfamily
  2110.01635}}.

\bibitem{Sun:2020sgn}
Z.~Sun, \emph{{Higher spin de Sitter quasinormal modes}},
  \href{https://arxiv.org/abs/2010.09684}{{\ttfamily 2010.09684}}.

\bibitem{Xiao2014}
X.~Xiao, \emph{Holographic representation of local operators in de sitter
  space}, \href{https://doi.org/10.1103/PhysRevD.90.024061}{\emph{Physical
  Review D - Particles, Fields, Gravitation and Cosmology} {\bfseries 90}
  (2014) 024061}.

\bibitem{Sleight:2021plv}
C.~Sleight and M.~Taronna, \emph{{From dS to AdS and back}},
  \href{https://arxiv.org/abs/2109.02725}{{\ttfamily 2109.02725}}.

\bibitem{Garidi:2003bg}
T.~Garidi, J.P.~Gazeau and M.V.~Takook, \emph{{'Massive' spin two field in de
  Sitter space}}, \href{https://doi.org/10.1063/1.1599055}{\emph{J. Math.
  Phys.} {\bfseries 44} (2003) 3838}
  [\href{https://arxiv.org/abs/hep-th/0302022}{{\ttfamily hep-th/0302022}}].

\bibitem{Takook:2014paa}
M.V.~Takook, \emph{{Quantum Field Theory in de Sitter Universe: Ambient Space
  Formalism}},  \href{https://arxiv.org/abs/1403.1204}{{\ttfamily 1403.1204}}.

\bibitem{Fronsdal:1978vb}
C.~Fronsdal, \emph{{Singletons and Massless, Integral Spin Fields on de Sitter
  Space (Elementary Particles in a Curved Space. 7.}},
  \href{https://doi.org/10.1103/PhysRevD.20.848}{\emph{Phys. Rev. D} {\bfseries
  20} (1979) 848}.

\bibitem{Fang:1979hq}
J.~Fang and C.~Fronsdal, \emph{{Massless, Half Integer Spin Fields in De Sitter
  Space}}, \href{https://doi.org/10.1103/PhysRevD.22.1361}{\emph{Phys. Rev. D}
  {\bfseries 22} (1980) 1361}.

\bibitem{Huguet:2006fe}
E.~Huguet, J.~Queva and J.~Renaud, \emph{{Conformally related massless fields
  in dS, AdS and Minkowski spaces}},
  \href{https://doi.org/10.1103/PhysRevD.73.084025}{\emph{Phys. Rev. D}
  {\bfseries 73} (2006) 084025}
  [\href{https://arxiv.org/abs/gr-qc/0603031}{{\ttfamily gr-qc/0603031}}].

\bibitem{Faci:2009un}
S.~Faci, E.~Huguet, J.~Queva and J.~Renaud, \emph{{Conformally covariant
  quantization of Maxwell field in de Sitter space}},
  \href{https://doi.org/10.1103/PhysRevD.80.124005}{\emph{Phys. Rev. D}
  {\bfseries 80} (2009) 124005}
  [\href{https://arxiv.org/abs/0910.1279}{{\ttfamily 0910.1279}}].

\bibitem{Bros:1995js}
J.~Bros and U.~Moschella, \emph{{Two point functions and quantum fields in de
  Sitter universe}},
  \href{https://doi.org/10.1142/S0129055X96000123}{\emph{Rev. Math. Phys.}
  {\bfseries 8} (1996) 327}
  [\href{https://arxiv.org/abs/gr-qc/9511019}{{\ttfamily gr-qc/9511019}}].

\bibitem{Bros:1994dn}
J.~Bros, U.~Moschella and J.P.~Gazeau, \emph{{Quantum field theory in the de
  Sitter universe}},
  \href{https://doi.org/10.1103/PhysRevLett.73.1746}{\emph{Phys. Rev. Lett.}
  {\bfseries 73} (1994) 1746}.

\bibitem{Henningson:1998cd}
M.~Henningson and K.~Sfetsos, \emph{{Spinors and the AdS / CFT
  correspondence}},
  \href{https://doi.org/10.1016/S0370-2693(98)00559-0}{\emph{Phys. Lett. B}
  {\bfseries 431} (1998) 63}
  [\href{https://arxiv.org/abs/hep-th/9803251}{{\ttfamily hep-th/9803251}}].

\bibitem{Candelas1975}
P.~Candelas and D.J.~Raine, \emph{General-relativistic quantum field theory: An
  exactly soluble model},
  \href{https://doi.org/10.1103/PhysRevD.12.965}{\emph{Phys. Rev. D} {\bfseries
  12} (1975) 965}.

\bibitem{Koksma2009}
J.F.~Koksma and T.~Prokopec, \emph{The fermion propagator in cosmological
  spaces with constant deceleration},
  \href{https://doi.org/10.1088/0264-9381/26/12/125003}{\emph{CQGra} {\bfseries
  26} (2009) 125003}.

\bibitem{Cotaescu2002}
I.I.~Cot{\u a}escu, \emph{Polarized dirac fermions in de sitter spacetime},
  \href{https://doi.org/10.1103/PhysRevD.65.084008}{\emph{Physical Review D}
  {\bfseries 65} (2002) 084008}.

\bibitem{Cotaescu2018}
I.I.~Cotaescu, \emph{Integral representation of the feynman propagators of the
  dirac fermions on the de sitter expanding universe},
  \href{https://doi.org/10.1140/epjc/s10052-018-6258-2}{\emph{European Physical
  Journal C} {\bfseries 78} (2018) }.

\bibitem{Camporesi1992}
R.~Camporesi, \emph{The spinor heat kernel in maximally symmetric spaces},
  \href{https://doi.org/10.1007/BF02100862}{\emph{Commun.Math.Phys.} {\bfseries
  148} (1992) 283}.

\bibitem{Camporesi1996}
R.~Camporesi and A.~Higuchi, \emph{On the eigen functions of the dirac operator
  on spheres and real hyperbolic spaces},
  \href{https://doi.org/10.1016/0393-0440(95)00042-9}{\emph{J.Geom.Phys.}
  {\bfseries 20} (1996) 1}.

\bibitem{Hertog:2019uhy}
T.~Hertog, G.~Tartaglino-Mazzucchelli and G.~Venken, \emph{{Spinors in
  Supersymmetric dS/CFT}},
  \href{https://doi.org/10.1007/JHEP10(2019)117}{\emph{JHEP} {\bfseries 10}
  (2019) 117} [\href{https://arxiv.org/abs/1905.01322}{{\ttfamily
  1905.01322}}].

\bibitem{Kawano:1999au}
T.~Kawano and K.~Okuyama, \emph{{Spinor exchange in AdS(d+1)}},
  \href{https://doi.org/10.1016/S0550-3213(99)00639-2}{\emph{Nucl. Phys. B}
  {\bfseries 565} (2000) 427}
  [\href{https://arxiv.org/abs/hep-th/9905130}{{\ttfamily hep-th/9905130}}].

\bibitem{Iliesiu:2015qra}
L.~Iliesiu, F.~Kos, D.~Poland, S.S.~Pufu, D.~Simmons-Duffin and R.~Yacoby,
  \emph{{Bootstrapping 3D Fermions}},
  \href{https://doi.org/10.1007/JHEP03(2016)120}{\emph{JHEP} {\bfseries 03}
  (2016) 120} [\href{https://arxiv.org/abs/1508.00012}{{\ttfamily
  1508.00012}}].

\bibitem{Weinberg2005}
S.~Weinberg, \emph{Quantum contributions to cosmological correlations},
  \href{https://doi.org/10.1103/PHYSREVD.72.043514}{\emph{Physical Review D -
  Particles, Fields, Gravitation and Cosmology} {\bfseries 72} (2005) 1}.

\bibitem{Higuchi:2010xt}
A.~Higuchi, D.~Marolf and I.A.~Morrison, \emph{{On the Equivalence between
  Euclidean and In-In Formalisms in de Sitter QFT}},
  \href{https://doi.org/10.1103/PhysRevD.83.084029}{\emph{Phys. Rev. D}
  {\bfseries 83} (2011) 084029}
  [\href{https://arxiv.org/abs/1012.3415}{{\ttfamily 1012.3415}}].

\bibitem{Gorbenko:2019rza}
V.~Gorbenko and L.~Senatore, \emph{{$\lambda \phi^4$ in dS}},
  \href{https://arxiv.org/abs/1911.00022}{{\ttfamily 1911.00022}}.

\bibitem{Sleight:2017fpc}
C.~Sleight and M.~Taronna, \emph{{Spinning Witten Diagrams}},
  \href{https://doi.org/10.1007/JHEP06(2017)100}{\emph{JHEP} {\bfseries 06}
  (2017) 100} [\href{https://arxiv.org/abs/1702.08619}{{\ttfamily
  1702.08619}}].

\bibitem{Sleight:2019hfp}
C.~Sleight and M.~Taronna, \emph{{Bootstrapping Inflationary Correlators in
  Mellin Space}}, \href{https://doi.org/10.1007/JHEP02(2020)098}{\emph{JHEP}
  {\bfseries 02} (2020) 098}
  [\href{https://arxiv.org/abs/1907.01143}{{\ttfamily 1907.01143}}].

\bibitem{Sleight:2019mgd}
C.~Sleight, \emph{{A Mellin Space Approach to Cosmological Correlators}},
  \href{https://doi.org/10.1007/JHEP01(2020)090}{\emph{JHEP} {\bfseries 01}
  (2020) 090} [\href{https://arxiv.org/abs/1906.12302}{{\ttfamily
  1906.12302}}].

\bibitem{Sleight:2020obc}
C.~Sleight and M.~Taronna, \emph{{From AdS to dS Exchanges: Spectral
  Representation, Mellin Amplitudes and Crossing}},
  \href{https://arxiv.org/abs/2007.09993}{{\ttfamily 2007.09993}}.

\bibitem{DiPietro:2021sjt}
L.~Di~Pietro, V.~Gorbenko and S.~Komatsu, \emph{{Analyticity and Unitarity for
  Cosmological Correlators}},
  \href{https://arxiv.org/abs/2108.01695}{{\ttfamily 2108.01695}}.

\bibitem{maldacena2003}
J.~Maldacena, \emph{Non-gaussian features of primordial fluctuations in single
  field inflationary models},
  \href{https://doi.org/10.1088/1126-6708/2003/05/013}{\emph{Journal of High
  Energy Physics} {\bfseries 2003} (2003) 013}.

\bibitem{Bzowski2013}
A.~Bzowski, P.~McFadden and K.~Skenderis, \emph{Holography for inflation using
  conformal perturbation theory},
  \href{https://doi.org/10.1007/JHEP04(2013)047}{\emph{Journal of High Energy
  Physics} {\bfseries 2013} (2013) 47}.

\bibitem{McFadden2010b}
P.~McFadden and K.~Skenderis, \emph{Holography for cosmology},
  \href{https://doi.org/10.1103/PHYSREVD.81.021301}{\emph{Physical Review D -
  Particles, Fields, Gravitation and Cosmology} {\bfseries 81} (2010) }.

\bibitem{Pimentel2014}
G.L.~Pimentel, \emph{Inflationary consistency conditions from a wavefunctional
  perspective}, \href{https://doi.org/10.1007/JHEP02(2014)124}{\emph{JHEP}
  {\bfseries 02} (2014) 124}.

\bibitem{Anninos2014}
D.~Anninos, T.~Anous, D.Z.~Freedman and G.~Konstantinidis, \emph{Late-time
  structure of the bunch-davies de sitter wavefunction},
  \href{https://doi.org/10.1088/1475-7516/2015/11/048}{\emph{Journal of
  Cosmology and Astroparticle Physics} {\bfseries 2015} (2014) }.

\bibitem{Mata2013}
I.~Mata, S.~Raju and S.P.~Trivedi, \emph{Cmb from cft},
  \href{https://doi.org/10.1007/JHEP07(2013)015}{\emph{Journal of High Energy
  Physics} {\bfseries 2013} (2013) 15}.

\bibitem{Harlow:2011ke}
D.~Harlow and D.~Stanford, \emph{{Operator Dictionaries and Wave Functions in
  AdS/CFT and dS/CFT}},  \href{https://arxiv.org/abs/1104.2621}{{\ttfamily
  1104.2621}}.

\bibitem{Hogervorst:2021uvp}
M.~Hogervorst, J.a.~Penedones and K.S.~Vaziri, \emph{{Towards the
  non-perturbative cosmological bootstrap}},
  \href{https://arxiv.org/abs/2107.13871}{{\ttfamily 2107.13871}}.

\bibitem{Baumann2018}
D.~Baumann, G.~Goon, H.~Lee and G.L.~Pimentel, \emph{Partially massless fields
  during inflation},
  \href{https://doi.org/10.1007/JHEP04(2018)140}{\emph{Journal of High Energy
  Physics 2018 2018:4} {\bfseries 2018} (2018) 1}.

\bibitem{Baumann:2019oyu}
D.~Baumann, C.~Duaso~Pueyo, A.~Joyce, H.~Lee and G.L.~Pimentel, \emph{{The
  cosmological bootstrap: weight-shifting operators and scalar seeds}},
  \href{https://doi.org/10.1007/JHEP12(2020)204}{\emph{JHEP} {\bfseries 12}
  (2020) 204} [\href{https://arxiv.org/abs/1910.14051}{{\ttfamily
  1910.14051}}].

\bibitem{Baumann:2020dch}
D.~Baumann, C.~Duaso~Pueyo, A.~Joyce, H.~Lee and G.L.~Pimentel, \emph{{The
  Cosmological Bootstrap: Spinning Correlators from Symmetries and
  Factorization}},  \href{https://arxiv.org/abs/2005.04234}{{\ttfamily
  2005.04234}}.

\bibitem{Baumann:2020ksv}
D.~Baumann, C.~Duaso~Pueyo and A.~Joyce, \emph{{Bootstrapping Cosmological
  Correlations}},
  \href{https://doi.org/10.22661/AAPPSBL.2020.30.6.02}{\emph{AAPPS Bull.}
  {\bfseries 30} (2020) 2}.

\bibitem{Arkani-Hamed:2018kmz}
N.~Arkani-Hamed, D.~Baumann, H.~Lee and G.L.~Pimentel, \emph{{The Cosmological
  Bootstrap: Inflationary Correlators from Symmetries and Singularities}},
  \href{https://doi.org/10.1007/JHEP04(2020)105}{\emph{JHEP} {\bfseries 04}
  (2020) 105} [\href{https://arxiv.org/abs/1811.00024}{{\ttfamily
  1811.00024}}].

\bibitem{Arkani-Hamed:2017fdk}
N.~Arkani-Hamed, P.~Benincasa and A.~Postnikov, \emph{{Cosmological Polytopes
  and the Wavefunction of the Universe}},
  \href{https://arxiv.org/abs/1709.02813}{{\ttfamily 1709.02813}}.

\bibitem{Goodhew:2021oqg}
H.~Goodhew, S.~Jazayeri, M.H.~Gordon~Lee and E.~Pajer, \emph{{Cutting
  cosmological correlators}},
  \href{https://doi.org/10.1088/1475-7516/2021/08/003}{\emph{JCAP} {\bfseries
  08} (2021) 003} [\href{https://arxiv.org/abs/2104.06587}{{\ttfamily
  2104.06587}}].

\bibitem{Jazayeri2021}
S.~Jazayeri, E.~Pajer and D.~Stefanyszyn, \emph{From locality and unitarity to
  cosmological correlators},
  \href{https://doi.org/10.1007/JHEP10(2021)065}{\emph{JHEP} {\bfseries 10}
  (2021) 065}.

\bibitem{Bonifacio2021}
J.~Bonifacio, E.~Pajer and D.-G.~Wang, \emph{From amplitudes to contact
  cosmological correlators},
  \href{https://doi.org/10.1007/JHEP10(2021)001}{\emph{arXiv} {\bfseries 2021}
  (2021) arXiv:2106.15468}.

\bibitem{Isono:2017grm}
H.~Isono, \emph{{On conformal correlators and blocks with spinors in general
  dimensions}}, \href{https://doi.org/10.1103/PhysRevD.96.065011}{\emph{Phys.
  Rev. D} {\bfseries 96} (2017) 065011}
  [\href{https://arxiv.org/abs/1706.02835}{{\ttfamily 1706.02835}}].

\bibitem{bruhat:1982}
Y.~Choquet-Bruhat, C.~DeWitt-Morette and M.~Dillard-Bleick, \emph{Analysis,
  manifolds, and physics}, North-Holland Pub. Co. (1982).

\bibitem{Costa:2014rya}
M.S.~Costa and T.~Hansen, \emph{{Conformal correlators of mixed-symmetry
  tensors}}, \href{https://doi.org/10.1007/JHEP02(2015)151}{\emph{JHEP}
  {\bfseries 02} (2015) 151} [\href{https://arxiv.org/abs/1411.7351}{{\ttfamily
  1411.7351}}].

\bibitem{Curry:2014yoa}
S.~Curry and A.R.~Gover, \emph{{An introduction to conformal geometry and
  tractor calculus, with a view to applications in general relativity}},
  \href{https://arxiv.org/abs/1412.7559}{{\ttfamily 1412.7559}}.

\bibitem{macdonald2013hypergeometric}
I.G.~{Macdonald}, \emph{Hypergeometric functions i}, {\emph{arXiv e-prints}
  (2013) } [\href{https://arxiv.org/abs/1309.4568}{{\ttfamily 1309.4568}}].

\bibitem{Bousso:2001mw}
R.~Bousso, A.~Maloney and A.~Strominger, \emph{{Conformal vacua and entropy in
  de Sitter space}},
  \href{https://doi.org/10.1103/PhysRevD.65.104039}{\emph{Phys. Rev. D}
  {\bfseries 65} (2002) 104039}
  [\href{https://arxiv.org/abs/hep-th/0112218}{{\ttfamily hep-th/0112218}}].

\bibitem{Allen:1985ux}
B.~Allen, \emph{{Vacuum States in de Sitter Space}},
  \href{https://doi.org/10.1103/PhysRevD.32.3136}{\emph{Phys. Rev. D}
  {\bfseries 32} (1985) 3136}.

\bibitem{Sasaki:1994yt}
M.~Sasaki, T.~Tanaka and K.~Yamamoto, \emph{{Euclidean vacuum mode functions
  for a scalar field on open de Sitter space}},
  \href{https://doi.org/10.1103/PhysRevD.51.2979}{\emph{Phys. Rev. D}
  {\bfseries 51} (1995) 2979}
  [\href{https://arxiv.org/abs/gr-qc/9412025}{{\ttfamily gr-qc/9412025}}].

\bibitem{Strominger:2001pn}
A.~Strominger, \emph{{The dS / CFT correspondence}},
  \href{https://doi.org/10.1088/1126-6708/2001/10/034}{\emph{JHEP} {\bfseries
  10} (2001) 034} [\href{https://arxiv.org/abs/hep-th/0106113}{{\ttfamily
  hep-th/0106113}}].

\bibitem{Adshead2009}
P.~Adshead, R.~Easther and E.A.~Lim, \emph{``in-in'' formalism and cosmological
  perturbations},
  \href{https://doi.org/10.1103/physrevd.80.083521}{\emph{Physical Review D}
  {\bfseries 80} (2009) }.

\bibitem{Trautman:1992}
A.~Trautman, \emph{Spinors and the dirac operator on hypersurfaces. i. general
  theory}, \href{https://doi.org/10.1063/1.529852}{\emph{Journal of
  Mathematical Physics} {\bfseries 33} (1992) 4011}.

\bibitem{Trautman:1995fr}
A.~Trautman, \emph{{The Dirac operator on hypersurfaces}}, {\emph{Acta Phys.
  Polon. B} {\bfseries 26} (1995) 1283}
  [\href{https://arxiv.org/abs/hep-th/9810018}{{\ttfamily hep-th/9810018}}].

\bibitem{Anninos:2017eib}
D.~Anninos, F.~Denef, R.~Monten and Z.~Sun, \emph{{Higher Spin de Sitter
  Hilbert Space}}, \href{https://doi.org/10.1007/JHEP10(2019)071}{\emph{JHEP}
  {\bfseries 10} (2019) 071}
  [\href{https://arxiv.org/abs/1711.10037}{{\ttfamily 1711.10037}}].

\bibitem{Deser2001}
S.~Deser and A.~Waldron, \emph{Partial masslessness of higher spins in (a)ds},
  \href{https://doi.org/10.1016/S0550-3213(01)00212-7}{\emph{Nuclear Physics B}
  {\bfseries 607} (2001) 577}.

\bibitem{Deser:2003gw}
S.~Deser and A.~Waldron, \emph{{Arbitrary spin representations in de Sitter
  from dS / CFT with applications to dS supergravity}},
  \href{https://doi.org/10.1016/S0550-3213(03)00348-1}{\emph{Nucl. Phys. B}
  {\bfseries 662} (2003) 379}
  [\href{https://arxiv.org/abs/hep-th/0301068}{{\ttfamily hep-th/0301068}}].

\bibitem{Karateev:2018uk}
D.~Karateev, P.~Kravchuk and D.~Simmons-Duffin, \emph{Weight {Shifting}
  {Operators} and {Conformal} {Blocks}},
  \href{https://doi.org/10.1007/JHEP02(2018)081}{\emph{Journal of High Energy
  Physics} {\bfseries 2018} (2018) 81}.

\bibitem{Sachs:1962zza}
R.~Sachs, \emph{{Asymptotic symmetries in gravitational theory}},
  \href{https://doi.org/10.1103/PhysRev.128.2851}{\emph{Phys. Rev.} {\bfseries
  128} (1962) 2851}.

\bibitem{Arnowitt:1962hi}
R.L.~Arnowitt, S.~Deser and C.W.~Misner, \emph{{The Dynamics of general
  relativity}}, \href{https://doi.org/10.1007/s10714-008-0661-1}{\emph{Gen.
  Rel. Grav.} {\bfseries 40} (2008) 1997}
  [\href{https://arxiv.org/abs/gr-qc/0405109}{{\ttfamily gr-qc/0405109}}].

\bibitem{Anninos:2010zf}
D.~Anninos, G.S.~Ng and A.~Strominger, \emph{{Asymptotic Symmetries and Charges
  in De Sitter Space}},
  \href{https://doi.org/10.1088/0264-9381/28/17/175019}{\emph{Class. Quant.
  Grav.} {\bfseries 28} (2011) 175019}
  [\href{https://arxiv.org/abs/1009.4730}{{\ttfamily 1009.4730}}].

\bibitem{Anninos:2011jp}
D.~Anninos, G.S.~Ng and A.~Strominger, \emph{{Future Boundary Conditions in De
  Sitter Space}}, \href{https://doi.org/10.1007/JHEP02(2012)032}{\emph{JHEP}
  {\bfseries 02} (2012) 032} [\href{https://arxiv.org/abs/1106.1175}{{\ttfamily
  1106.1175}}].

\bibitem{Visser:2017atf}
M.~Visser, \emph{{How to Wick rotate generic curved spacetime}},
  \href{https://arxiv.org/abs/1702.05572}{{\ttfamily 1702.05572}}.

\bibitem{Schlingemann:1998cw}
D.~Schlingemann, \emph{{From Euclidean field theory to quantum field theory}},
  \href{https://doi.org/10.1142/S0129055X99000362}{\emph{Rev. Math. Phys.}
  {\bfseries 11} (1999) 1151}
  [\href{https://arxiv.org/abs/hep-th/9802035}{{\ttfamily hep-th/9802035}}].

\bibitem{Schlingemann:1999mk}
D.~Schlingemann, \emph{{Euclidean field theory on a sphere}},
  \href{https://arxiv.org/abs/hep-th/9912235}{{\ttfamily hep-th/9912235}}.

\bibitem{David:2021wrw}
J.R.~David and J.~Mukherjee, \emph{{Partition functions of $p$-forms from
  Harish-Chandra characters}},
  \href{https://arxiv.org/abs/2105.03662}{{\ttfamily 2105.03662}}.

\bibitem{anninos2012}
D.~Anninos, S.A.~Hartnoll and D.M.~Hofman, \emph{Static patch solipsism:
  conformal symmetry of the de sitter worldline},
  \href{https://doi.org/10.1088/0264-9381/29/7/075002}{\emph{Classical and
  Quantum Gravity} {\bfseries 29} (2012) 075002}.

\bibitem{Banks:2006rx}
T.~Banks, B.~Fiol and A.~Morisse, \emph{{Towards a quantum theory of de Sitter
  space}}, \href{https://doi.org/10.1088/1126-6708/2006/12/004}{\emph{JHEP}
  {\bfseries 12} (2006) 004}
  [\href{https://arxiv.org/abs/hep-th/0609062}{{\ttfamily hep-th/0609062}}].

\bibitem{Law:2020cpj}
Y.T.A.~Law, \emph{{A compendium of sphere path integrals}},
  \href{https://doi.org/10.1007/JHEP12(2021)213}{\emph{JHEP} {\bfseries 21}
  (2020) 213} [\href{https://arxiv.org/abs/2012.06345}{{\ttfamily
  2012.06345}}].

\bibitem{Muhlmann}
B.~Mühlmann, \emph{The two-sphere partition function in two-dimensional
  quantum gravity at fixed area},
  \href{https://doi.org/10.1007/JHEP09(2021)189}{\emph{JHEP} {\bfseries 09}
  (2021) }.

\bibitem{Anninos2021}
D.~Anninos, T.~Bautista and B.~Mühlmann, \emph{The two-sphere partition
  function in two-dimensional quantum gravity},
  \href{https://doi.org/10.1007/JHEP09(2021)116}{\emph{JHEP} {\bfseries 09}
  (2021) 116}.

\bibitem{anninos2019Holo}
D.~Anninos, D.A.~Galante and D.M.~Hofman, \emph{De sitter horizons \&
  holographic liquids},
  \href{https://doi.org/10.1007/jhep07(2019)038}{\emph{Journal of High Energy
  Physics} {\bfseries 2019} (2019) }.

\bibitem{Harris2021}
D.~Anninos and E.~Harris, \emph{Three-dimensional de sitter horizon
  thermodynamics}, \href{https://doi.org/10.1007/JHEP10(2021)091}{\emph{Journal
  of High Energy Physics 2021 2021:10} {\bfseries 2021} (2021) 1}.

\bibitem{Anninos:2020hfj}
D.~Anninos, F.~Denef, Y.T.A.~Law and Z.~Sun, \emph{{Quantum de Sitter horizon
  entropy from quasicanonical bulk, edge, sphere and topological string
  partition functions}},
  \href{https://doi.org/10.1007/JHEP01(2022)088}{\emph{JHEP} {\bfseries 01}
  (2022) 088} [\href{https://arxiv.org/abs/2009.12464}{{\ttfamily
  2009.12464}}].

\bibitem{herzog2017}
C.P.~Herzog and K.-W.~Huang, \emph{Boundary conformal field theory and a
  boundary central charge},
  \href{https://doi.org/10.1007/jhep10(2017)189}{\emph{Journal of High Energy
  Physics} {\bfseries 2017} (2017) }.

\bibitem{Herzog:2021spv}
C.P.~Herzog and V.~Schaub, \emph{{A Sum Rule for Boundary Contributions to the
  Trace Anomaly}},  \href{https://arxiv.org/abs/2107.11604}{{\ttfamily
  2107.11604}}.

\bibitem{David2019}
A.~David, N.~Fischer and Y.~Neiman, \emph{Spinor-helicity variables for
  cosmological horizons in de sitter space},
  \href{https://doi.org/10.1103/PHYSREVD.100.045005}{\emph{Phys.Rev.D}
  {\bfseries 100} (2019) }.

\bibitem{Caron-Huot:2021kjy}
S.~Caron-Huot and Y.-Z.~Li, \emph{{Helicity basis for three-dimensional
  conformal field theory}},  \href{https://arxiv.org/abs/2102.08160}{{\ttfamily
  2102.08160}}.

\bibitem{Weinberg:1995mt}
S.~Weinberg, \emph{{The Quantum theory of fields. Vol. 1: Foundations}},
  Cambridge University Press (6, 2005).

\bibitem{Anninos:2011ui}
D.~Anninos, T.~Hartman and A.~Strominger, \emph{{Higher Spin Realization of the
  dS/CFT Correspondence}},
  \href{https://doi.org/10.1088/1361-6382/34/1/015009}{\emph{Class. Quant.
  Grav.} {\bfseries 34} (2017) 015009}
  [\href{https://arxiv.org/abs/1108.5735}{{\ttfamily 1108.5735}}].

\bibitem{Witten:2001kn}
E.~Witten, \emph{{Quantum gravity in de Sitter space}},  in \emph{{Strings
  2001: International Conference}}, 6, 2001
  [\href{https://arxiv.org/abs/hep-th/0106109}{{\ttfamily hep-th/0106109}}].

\bibitem{Dobrev:1977qv}
V.K.~Dobrev, G.~Mack, V.B.~Petkova, S.G.~Petrova and I.T.~Todorov,
  \emph{{Harmonic Analysis on the n-Dimensional Lorentz Group and Its
  Application to Conformal Quantum Field Theory}}, vol.~63, Springer (1977),
  \href{https://doi.org/10.1007/BFb0009678}{10.1007/BFb0009678}.

\bibitem{Sun2021}
Z.~Sun, \emph{A note on the representations of $\text\{SO\}(1,d+1)$},
  {\emph{arXiv} (2021) arXiv:2111.04591}.

\bibitem{Thieleker1973}
E.~Thieleker, \emph{On the quasi-simple irreducible representations of the
  lorentz groups}, \href{https://doi.org/10.2307/1996515}{\emph{Transactions of
  the American Mathematical Society} {\bfseries 179} (1973) 465}.

\bibitem{Thieleker1974}
E.A.~Thieleker, \emph{The unitary representations of the generalized lorentz
  groups}, \href{https://doi.org/10.2307/1996891}{\emph{Transactions of the
  American Mathematical Society} {\bfseries 199} (1974) 327}.

\bibitem{Joung2006}
E.~Joung, J.~Mourad and R.~Parentani, \emph{Group theoretical approach to
  quantum fields in de sitter space, i. the principal series},
  \href{https://doi.org/10.1088/1126-6708/2006/08/082}{\emph{Journal of High
  Energy Physics} {\bfseries 2006} (2006) }.

\bibitem{Joung2007}
E.~Joung, J.~Mourad and R.~Parentani, \emph{Group theoretical approach to
  quantum fields in de sitter space ii. the complementary and discrete series},
  \href{https://doi.org/10.1088/1126-6708/2007/09/030}{\emph{JHEP} {\bfseries
  2007} (2007) 030}.

\bibitem{Anous2020}
T.~Anous and J.~Skulte, \emph{An invitation to the principal series},
  \href{https://doi.org/10.21468/scipostphys.9.3.028}{\emph{SciPost Physics}
  {\bfseries 9} (2020) }.

\bibitem{Newton1950}
T.D.~Newton, \emph{A note on the representations of the de sitter group},
  \href{https://doi.org/10.2307/1969376}{\emph{The Annals of Mathematics}
  {\bfseries 51} (1950) 730}.

\bibitem{Basile2017}
T.~Basile, X.~Bekaert, N.~Boulanger, T.~Basile, X.~Bekaert and N.~Boulanger,
  \emph{Mixed-symmetry fields in de sitter space: a group theoretical glance},
  \href{https://doi.org/10.1007/JHEP05(2017)081}{\emph{JHEP} {\bfseries 2017}
  (2017) 81}.

\bibitem{Simmons-Duffin:2014wb}
D.~Simmons-Duffin, \emph{Projectors, {Shadows}, and {Conformal} {Blocks}},
  \href{https://doi.org/10.1007/JHEP04(2014)146}{\emph{Journal of High Energy
  Physics} {\bfseries 2014} (2014) 146}.

\bibitem{Joung2016}
E.~Joung and K.~Mkrtchyan, \emph{Partially-massless higher-spin algebras and
  their finite-dimensional truncations},
  \href{https://doi.org/10.1007/JHEP01(2016)003}{\emph{Journal of High Energy
  Physics} {\bfseries 2016} (2016) 1}.

\bibitem{VanProeyen:1999ni}
A.~Van~Proeyen, \emph{{Tools for supersymmetry}}, {\emph{Ann. U. Craiova Phys.}
  {\bfseries 9} (1999) 1}
  [\href{https://arxiv.org/abs/hep-th/9910030}{{\ttfamily hep-th/9910030}}].

\bibitem{Stahl2016}
C.~Stahl, E.~Strobel and S.S.~Xue, \emph{Fermionic current and schwinger effect
  in de sitter spacetime},
  \href{https://doi.org/10.1103/PHYSREVD.93.025004}{\emph{Physical Review D}
  {\bfseries 93} (2016) }.

\bibitem{Higuchi:1985ad}
A.~Higuchi, \emph{Symmetric tensor fields in de sitter space-time},
  {\emph{YTP-85-22} (1985) }.

\bibitem{Higuchi1987}
A.~Higuchi, \emph{Forbidden mass range for spin-2 field theory in de sitter
  space-time},
  \href{https://doi.org/10.1016/0550-3213(87)90691-2}{\emph{Nucl.Phys.B}
  {\bfseries 282} (1987) 397}.

\bibitem{Higuchi1987cf}
A.~Higuchi, \emph{Symmetric tensor spherical harmonics on the $n$ sphere and
  their application to the de sitter group so($n$,1)},
  \href{https://doi.org/10.1063/1.527513}{\emph{J.Math.Phys.} {\bfseries 28}
  (1987) 1553}.

\end{thebibliography}\endgroup

\end{document}